\documentclass[11pt]{article}
\usepackage[margin=1in]{geometry}
\usepackage{url}
\usepackage{numprint}
\usepackage{setspace}                % See geometry.pdf to learn the layout options. There are lots.
\usepackage{multirow}
\usepackage{graphicx}
\usepackage[numbers]{natbib}
\usepackage{amssymb,amsmath,bbm,float,amsthm}
\usepackage{epstopdf,dsfont}
\usepackage{xcolor}
\usepackage{enumerate}
\usepackage{prodint}
\usepackage{csquotes}

\usepackage{xr}%for external documents

\DeclareMathOperator*{\argmax}{arg\,max}

\newtheorem{theorem}{Theorem}
\newtheorem{proposition}{Proposition}

\makeatletter
\newcommand{\addresseshere}{%
	\enddoc@text\let\enddoc@text\relax
}
\makeatother

\begin{document}

% Title of paper
\title{Change-point detection in regression models for ordered data via the max-EM algorithm}%based on% with statistical breakpoint tests}
\author{Modibo Diabaté$^{1}$, Grégory Nuel$^{2}$ and Olivier Bouaziz$^{1}$}
\date{$^1$Université Paris Cité, CNRS, MAP5, F-75006 Paris, France\\
	$^2$LPSM (UMR CNRS 8001), Sorbonne Universit\'e, France
%	$^3$ISEN Yncréa Méditerranée, France 
}
%\author{Modibo Diabaté$^{1}$}
%\address{Modibo Diabaté,
%	ISEN Yncréa Méditerranée, France \& MAP5 (UMR 8145), Université de Paris, France}
%\email{modibo.diabate@yncrea.fr}
%
%\author{Olivier Bouaziz$^{2}$}
%\address{Olivier Bouaziz,
%	MAP5 (UMR CNRS 8145), Université de Paris, France}
%\email{olivier.bouaziz@parisdescartes.fr}
%
%\author{Grégory Nuel$^{3}$}
%\address{Grégory Nuel,
%	LPSM (UMR CNRS 8001), Sorbonne Université, France}
%\email{nuel@math.cnrs.fr}

\maketitle

\begin{abstract}

We consider the problem of breakpoint detection in a regression modeling framework. To that end, we introduce a novel method, the max-EM algorithm which combines a constrained Hidden Markov Model with the Classification-EM (CEM) algorithm. This algorithm has linear complexity and provides accurate breakpoints detection and parameter estimations. We derive a theoretical result that shows that the likelihood of the data as a function of the regression parameters and the breakpoints location is increased at each step of the algorithm. We also present two initialization methods for the location of the breakpoints in order to deal with local maxima issues. Finally, a statistical test in the one breakpoint situation is developed. Simulation experiments based on linear, logistic, Poisson and Accelerated Failure Time regression models show that the final method that includes the initialization procedure and the max-EM algorithm has a strong performance both in terms of parameters estimation and breakpoints detection. The statistical test is also evaluated and exhibits a correct rejection rate under the null hypothesis and a strong power under various alternatives. Two real dataset are analyzed, the UCI bike sharing and the health disease data, where the interest of the method to detect heterogeneity in the distribution of the data is illustrated.

\end{abstract}


\section{Introduction}

\label{sec:Intro}

Breakpoint modeling is a major topic in many applications and taking them into account generally allows a better understanding of the studied problem. In finance, the detection of points of variation in time series of volatility of financial assets allows a better appreciation of the market risks and represents a subject of great interest \citep[see][]{shiller1992market, gallant1997estimation, cho2012multiscale}. Other examples include environmental changes over time~\citep[see][]{killick2012optimal, reeves2007review, killick2010detection} or speech perception in the analysis of sound signals~\citep[see][]{gomot2006change, davis2007hearing}. It is also an important and relevant topic in many medical applications, as the study of breakpoint detection allows to detect heterogeneity in patients data: this is particularly interesting in personalized medicine where the goal is to optimize treatment strategies. Applications of breakpoint models also include genomic data in cancer studies like in~\cite{venkatraman2007faster, shen2009integrative, venkatraman2007dnacopy, zhang2010detecting} where the efficient detection of the change in the number of DNA copies in cancer data makes it possible to detect the presence of cancer cells (characterized by a faster division frequency), or even to study the progression and type of a cancerous tumor. Several approaches have been proposed to deal with such problems and breakpoint detection methods can be separated in two main classes: exact breakpoint calculation and statistical methods. In the first case, the aim is to develop an efficient algorithm that exhaustively explores all possible segmentations (corresponding to all possible breakpoints) while in the second case, the aim is to build a statistical model that aims at finding the most probable segmentation.

Exact calculation of breakpoints can be performed using dynamic programming with the Optimal Partitioning (OP) approach~\citep[see][]{jackson2005algorithm}. However, this method has a high computational complexity of order $O(n^2)$ which makes it intractable to use with large datasets. Optimized versions of this dynamic algorithm involving a pruning step have been proposed to reduce the algorithmic complexity. In particular, the Pruned Exact Linear Time (PELT) method introduced by~\cite{killick2012optimal}  has a linear computational cost when the number of change-points increases as we observe more data. Many other algorithms have been introduced to attempt to reduce the time complexity of this algorithm. This is the case for instance of the Functional Pruning Optimal Partitioning (FPOP) algorithm~\citep[see][]{maidstone2017optimal, rigaill2019fpop} and its extension, the Generalized Functional Pruning Optimal Partitioning (GFPOP) algorithm~\citep[see][]{hocking2020constrained, runge2020gfpop}. These algorithms have the property that they can include constraints and they can consider a wide range of loss functions. See~\cite{runge2020gfpop} for a more detailed review of the dynamic programming based algorithms that were developed for breakpoint detection. However, all these methods are not suited to deal with regression modeling. They are tailored to the detection of breakpoints over a series of values of a response vector but they cannot include information from a covariate matrix. Also, in the simple mean model, where the differences in terms of segments is characterized by the mean of the response vector, the gfpop algorithm can only work under homoscedasticity. %~\citep[see][]{killick2012optimal}.

In this work, we present a general approach based on statistical models that extends the dynamic programming algorithms to regression modeling but is no longer based on exact breakpoints calculation. The main challenge is then to be able to extend the breakpoint detection to more general models while also keeping a good accuracy in breakpoint detection. In~\cite{bouaziz2018change} and~\cite{alarcon2019detecting}, the authors have proposed a methodology that combines Hidden Markov Model (HMM) methods and the Expectation maximization (EM) algorithm to achieve this goal, in a logistic and a Cox regression models, respectively. While the method has shown to be of interest to detect heterogeneity in binary or time to event data, it also suffers two major drawbacks. First, the algorithm is highly sensitive to the initialization value of the parameters, where several initialization choices may lead to different breakpoints and estimated parameters. Second, if the focus is mostly on breakpoint detection, the EM step is not adapted. This is because it makes a compromise by finding the most relevant regression parameters that maximize the averaged likelihood over all possible segmentations when the same value of the regression parameter is used in each segmentation. This lead us to the development of a new method, called the max-EM algorithm. In this method, the EM step is replaced by a Classification EM (CEM) step, inspired from the work of~\cite{celeux1992classification}. Moreover, the segments are modeled using HMM, as in~\cite{bouaziz2018change},  but we introduce a new forward-backward algorithm where the computation of the forward and backward quantities is performed by taking the maximum (instead of the sum) over a sequence of segments. We show that this new algorithm is well suited to the breakpoint detection problem where the aim is to find the best segmentation among a fix number of segments in a general regression framework. Then, we also present two strategies for the initialization of this iterative algorithm. The first one is based on the Fused Lasso (FL) method~\citep[see][]{tibshirani2005sparsity, rinaldo2009properties} where we implement the overparameterized setting with a number of segments equal to the number of individuals and we penalize the values of regression parameters over two consecutive segments. The second one is based on Binary Segmentation (BS) where the idea is to recursively apply the simple one breakpoint model~\citep[see][]{scott1974cluster}. Both approaches allow to derive a sequence of breakpoint candidates. From these, we run the max-EM algorithm for all possible combinations and keep the result from the model with the highest likelihood value. 
Finally, we address the problem of heterogeneity detection from a statistical point of view. More precisely, we develop a new statistical test in the one breakpoint situation. From a theoretical point of view, the derivation of the distribution of the statistical test is extremely difficult due to the fact that it involves the maximum over all possible segmentations of the maximum over all parameter values. This is why we derive asymptotic approximations of the likelihood ratio test from which the maximum over all possible segmentations can be easily computed. This provides a very useful and easily implementable statistical test. Our simulation results show that the max-EM algorithm works well in practice, both for the detection of breakpoints and the estimation of regression parameters. We observe that the initialization procedures find relevant breakpoints that allow to stabilize the results with an advantage over the BS initialization in terms of performance and computation time balance. Regarding the statistical test, we observed that it is well calibrated under the null hypothesis and has a strong power under various alternatives.

The paper is organized as follows. We first present the main goals of the paper in the next section. Then, the EM algorithm combined with HMM is recalled in Section~\ref{sec:methodoMaxEM}. We show in particular that it does not address the problem of breakpoint detection. We further introduce our new max-EM algorithm, we derive its theoretical properties and the two initialization procedures are presented. We conclude the section by presenting a standard Bayesian Information Criterion (BIC) used to select the number of breakpoints. 
In Section~\ref{sec:test}, we present the approximation formulas for the statistical test based on likelihood ratio computation. In Section~\ref{sec:simu} extensive simulation experiments are conducted: the performance of our method for breakpoints detection and parameters estimation is studied through several regression modeling (linear, logistic, Poisson and AFT regressions) and different number of breakpoints (from $1$ to $5$). The statistical test is also studied under the same regression models. In Section~\ref{sec:realdata} we study two real dataset using our new method: the UCI bike sharing dataset where the aim is to detect change of trends with respect to the date for the number of total daily rental bikes and the UCI heart disease dataset where the aim is to detect heterogeneity in the effect of fasting blood sugar on the risk of developing a heart disease. %we present similar results by using real data on xxx. The paper concludes with a discussion in Section 7.%statistical breakpoint tests using simulated data and considering different loss functions

\section{Objectives}

\label{sec:Objectives}

We consider a maximum likelihood based problem %method 
in the situation where the distribution of the data depends on $K$ segments. More specifically, we assume there exists $K-1$ breakpoints $(n_1^*,\ldots,n_{K-1}^*)\in\{1,\ldots,n-1\}$ such that $n_0^*=0<n_1^*<\cdots<n_{K-1}^*<n_K^*=n$ and for $k=1,\ldots,K$, $X_{n_{k-1}^*+1},\ldots,X_{n_{k}^*}$ are independent and identically distributed (iid) following a distribution with continuous/discrete probability distribution function, denoted $e_i(k;\theta^*_k)$, that depends on an unknown $d$ dimensional parameter $\theta^*_k\in \Theta\subset\mathbb R^d$. %density/probability
 %$\mathbb P(\cdot,\theta^*_k)$. This density/probability depends on an unknown $d$ dimensional parameter $\theta^*_k\in \Theta\subset\mathbb R^d$. 
Importantly, the number and location of the segments are also assumed to be unknown. Let $R_i\in \{1,\ldots,K\}$ be the latent variable representing the segment index associated to each individual: $R_i=k$ for $i\in\{n^*_{k-1}+1,\ldots,n^*_k\}$. Using this notation, $e_i(k;\theta_k)=\mathbb P(X_i\mid R_i=k;\theta_k)$ represents the conditional distribution of $X_i$ given $R_i=k$, evaluated at the parameter $\theta_k$. %In this work, the segment index is treated as an unknown parameter that we aim at estimating. 
For a given set of breakpoints and parameters, the log-likelihood of such a model can be written as:%We denote by $\mathbb P(\cdot\mid R=k;\theta^*_k)$ t
\begin{align}\label{eq:likelihood}
	\ell_n(\boldsymbol{\theta};n_{1:(K-1)})&=\log\big( \mathbb P(X_{1:n},R_{1:n}\mid\boldsymbol{\theta})\big)\nonumber\\%\prod_{i=1}^n%_{1:K}
	&=\sum_{k=1}^K \sum_{i\in \mathcal C_k} \log \big(e_i(k;\theta_k)\big)+\log\big(\mathbb P(R_{1:n})\big), 
	%\sum_{k=1}^K \sum_{i\in \mathcal C_k} \log\left(p_ke_i(k;\theta_k)\right),%\mathbb P(X_i\mid R_i=k;
\end{align}
where $\mathcal C_k=\{n_{k-1}+1,\ldots,n_{k}\}$ %$p_k=\mathbb P(R_i=k;\theta_k)$ 
and we use the compact notations $X_{1:n}$, $R_{1:n}$, $\boldsymbol{\theta}$, $n_{1:(K-1)}$ to represent the set of variables and parameters $X_1,\ldots,X_n$, $R_1,\ldots,R_n$, $\theta_1,\ldots,\theta_K$, $n_1,\ldots,n_{K-1}$ respectively. It should be noted that $\ell_n$ corresponds to the CML criterion ($C_1$) introduced in \cite{celeux1992classification}, since the $\mathbb P(R_{1:n})$ term can be omitted in the maximization. However, the major difference with their criterion comes from the structure of the $\mathcal C_k$ sets which can only contain ordered values of individuals in our case. In order to take into account this order of the individuals, we impose a Markov structure upon the $R_i$'s: we assume that each $R_i$ only depends on $R_{i-1}$, $i=2,\ldots,n$. 
%We also impose that $R_1=1$ and $R_n=K$.
We also impose that $\mathbb P(R_1=1)=1$ and we restrict our analysis to the set of Markov chains verifying $R_n=K$.

In practice, the interest of the method lies in the regression modeling of joint distributions, such that $X_i=(Y_i,Z_i)$, where $Y_i$ is an outcome variable and $Z_i$ a covariate vector of dimension $d$. Typically the conditional distribution of the $Y_i$'s given the $Z_i$'s will depend on $\theta_1^*,\ldots,\theta_K^*$ while the marginal distribution of the $Z_i$'s will be parameter free. In this regression framework, the conditional density of $X_i$ given $R_i=k$, $e_i(k;\theta_k)$, can be directly specified as following a regression model. In particular, in the simulation section, we consider the linear, the logistic, the Poisson and the Accelerated Failure Time (AFT) regression models.% for censored data.  

%where ${\color{orange}R} \in \{1,\ldots,K\}^n$, a constrained {\color{cyan}($R_n = K$)} hidden Markov process, represents a segmentation of $n$ points into $K$ segments %(no empty segments)
%(e.g: $R=1111222$ for $n=7$, $K=2$ and one breakpoint: $\mathrm{bp}=4$). Furthermore, we assume that the outcome variables $Y_i$ are generated according to a regression function in each segment. In the example of simple logistic regression model, we obtain:
%$$
%\mathbb{P}(Y_i=1|R_i=k, Z_i, \theta_k) = \frac{\text{e}^{Z_i\theta_k}}{1+\text{e}^{Z_i\theta_k}},
%\quad
%\mathbb{P}(Y_i=0|R_i=k, Z_i, \theta_k) = \frac{1}{1+\text{e}^{Z_i\theta_k}}
%$$

\subsection{First goal} 
The first goal of this paper is to develop a method for inferring the number and locations of the segments along with the estimation of the parameters $\theta_k$. 
This is done, when the number of breakpoints is fixed, by maximizing Equation~\eqref{eq:likelihood} with respect to both the $n_k$'s and $\theta_k$'s %(see section \ref{ssec:bruteforce})
\begin{align} \label{eq:likelihoodMaximization}
	\max_{n_1,\ldots,n_{K-1}} \sup_{\theta_1,\ldots,\theta_K}\ell_n(\boldsymbol{\theta};n_{1:(K-1)})%\mathcal C_k
\end{align}
This maximization problem can be directly solved sequentially by computing the maximum of $\ell_n(\theta_{1_K};n_{1:(K-1)})$ with respect to $\theta_k$ for each $\mathcal C_k$, and then by taking the maximum of all these values. This naive approach will be called the ``Brute force'' algorithm in the following. %e maximization problem in Equation~\eqref{eq:likelihoodMaximization}
It will accurately detect the breakpoints and the parameter values and is very simple to implement. However, the computation of our log-likelihood criterion for all possible segmentations is computationally very intensive ($O(n^{K+1})$ for the problem with $K$ breakpoints) and it is therefore not a feasible approach for large datasets or for several number of segments.%This method%as its name suggests, it is a naive approach that performs exhaustive calculations. T
%problem and $O(n^3)$ for a problem with only two breakpoints)}

As an alternative, one can use the EM algorithm to take into account the latent segment index. Models based on the EM algorithm and constrained Hidden Markov Model (HMM) were proposed in~\cite{bouaziz2018change} and~\cite{alarcon2019detecting}. Those methods are fast to execute (linear complexity) and provide high accuracy when properly initialized. However, we show in Section~\ref{sec:EM_standard} that the EM method does not solve the problem in Equation~\eqref{eq:likelihoodMaximization}. Instead, it attempts to find the $\boldsymbol{\theta}$ parameter that makes the best comprise when we average all possible segmentations and the same value of $\boldsymbol{\theta}$ is used in each segmentation.  
  This is why we introduce, in Section~\ref{sec:maxEM_method}, a novel method, called the max-EM algorithm, and show in Section~\ref{sec:theo_res_EM} that this max-EM algorithm is well adapted to the maximization problem of Equation~\eqref{eq:likelihoodMaximization} in the sense that each iteration of the algorithm is shown to increase the log-likelihood. As the algorithm is highly sensitive to parameters initialization, we also develop two different strategies for the initialization of the max-EM algorithm in Section~\ref{ssec:initialization}. %in~\cite{bouaziz2018change}%first%In this work,
%that extends the work from~\cite{alarcon2019detecting} which only considered binary segmentation for logistic regression models. % efficient
%
Since the max-EM algorithm only works for a fix value of $K$ we also propose, in Section \ref{ssec:bicInferNBbp}, an heuristic based on the Bayesian Information Criterion (BIC) to infer the number of breakpoints $K$. %is proposed 
The final max-EM algorithm, integrating the proposed initialization strategy, is implemented and evaluated on simulated data in Section~\ref{sec:simu}. %proportional hazard model
Various regression models and number of breakpoints are considered. 
%and regression models (linear, logistic, survival models) are considered. 
In the one breakpoint setting, our method is compared with the ``Brute force'' algorithm. In the absence of covariates, our approach is compared with the optimal dynamic programming algorithm GFPOP~\citep[see][]{runge2020gfpop} when a simple mean model is considered. All our results show that our method works well in practice and can extend the GFPOP method to regression modeling. %Finally, in Section~\ref{sec:realdata}, two real data are also analyzed using our method: the bike sharing dataset and the UCI heart disease dataset.

% is compared to the ``Brute force'' algorithm (in the one breakpoint setting) and to the optimal dynamic programming algorithm GFPOP \cite{runge2020gfpop} in {\color{black}Section~\ref{ssec:bpDetectionSimu} and Section~\ref{ssec:bpDetectionRealData}}.

\subsection{Second goal}\label{ssec:second}
The second goal of this paper is to develop a new statistical test in the one breakpoint scenario. In other words we propose a statistical test to make a decision between the two hypothesis
\begin{align}\label{eq:test}
	(H_0):\, &X_{1},\ldots,X_{n}\sim \mathcal L(\cdot,\theta^*) \text{ with }\theta^*\in \Theta \nonumber\\
	(H_1):\,&\exists \,n_1^*\in\{2,\ldots,n-1\} : X_{1},\ldots,X_{n_1^*}\sim \mathcal L(\cdot,\theta^*_1), X_{n_{1}^*+1},\ldots,X_{n}\sim \mathcal L(\cdot,\theta^*_2),\nonumber\\
	& \text{ with }\theta^*_1\neq\theta^*_2, (\theta^*_1,\theta^*_2)\in \Theta^2.
\end{align}
A likelihood based ratio test is presented in Section~\ref{sec:test} for this purpose. The statistical test requires to take the maximum of the log-likelihood ratio over all possible values of the breakpoint $n_1^*$, and for each value of $n_1^*$, over all possible values of the regression parameters. Deriving the exact or asymptotic distribution of this statistical test is extremely challenging. This is why we instead provide, in Section~\ref{sec:test}, an approximation formula of the log-likelihood ratio for any breakpoint value. The interest in this approximation formula lies in the fact that the $\boldsymbol{\theta}$ parameter and the Hessian matrix need only to be estimated under the null hypothesis. The score vector is also computed at the $\boldsymbol{\theta}$ parameter estimated under the null hypothesis but evaluated on the two segments. As a result, computing this approximated formula for all possible breakpoint values is extremely fast. In practice, this allows to easily test for breakpoint detections when using regression modeling. We also show in Section~\ref{ssec:bpDetectionSimu3} that the approximation formula works well on simulated data: under various regression models and breakpoint situations, we observe that using our formula the statistical test has the correct rejection rate under the null hypothesis and a good power under various alternative hypothesis.

\section{Breakpoint detection methodology}%: building of the max-EM algorithm}

\label{sec:methodoMaxEM}
%\section{Estimation methods: building of the max-EM algorithm}
%%\section{The max-EM algorithm}

% HMM \& EM algorithm for breakpoint computation
In this section, we present our approach based on the max-EM algorithm to perform breakpoint detection in ordered data. 
The max-EM approach is based on the use of a constrained HMM via a forward-backward type algorithm inspired from the EM algorithm. In Section~\ref{sec:EM_standard}, we first recall the EM method presented in~\cite{bouaziz2018change} and explain why this method does not maximize the criterion defined in Equation~\eqref{eq:likelihood}. We then introduce the max-EM algorithm in Section~\ref{sec:maxEM_method} and show in Section~\ref{sec:theo_res_EM} that each iteration of the algorithm increases the likelihood in Equation~\eqref{eq:likelihood}. In Section~\ref{ssec:initialization} we propose two different strategies for the initialization of the algorithm. In Section~\ref{ssec:bicInferNBbp}, we explain how the choice of the number of breakpoints $K$ can be done based on the Bayesian Information Criterion (BIC).% present the main aspects of these algorithms, then the derivation of the max-EM algorithm. 

\subsection{Review on the EM algorithm for ordered data in a HMM}\label{sec:EM_standard}% of the max-EM algorithm

The EM algorithm is an iterative method designed to maximize the observed likelihood $\mathbb{P}(X_{1:n} \mid\boldsymbol{\theta})$. Given a current parameter ${\boldsymbol{\theta}}^{(m)}$, the E-step is based on the computation of the quantity

\begin{align*}
\mathbb{Q}(\boldsymbol{\theta}\mid{\boldsymbol{\theta}}^{(m)}) = \mathbb{E} \left[\log \mathbb{P}(X_{1:n}, R_{1:n}\mid\theta)\mid X_{1:n};{\boldsymbol{\theta}}^{(m)}\right] &= \sum_{R_{1:n}} \mathbb{P}(R_{1:n}\mid X_{1:n}; {\boldsymbol{\theta}}^{(m)}) \log \mathbb{P}(X_{1:n}, R_{1:n}\mid\boldsymbol{\theta}),% Attention : l'ordre des sommes est invers� dans Celeux --> r�ponse Celeux regarde vraiment la vraisemblance dans l'intro alors qu'on est ici sur Q qui est diff�rent de la vraie vraisemblance.  
\end{align*}
where the sum is taken over all possible segmentations such that $R_n=K$. Introduce the weights 
\begin{align*}
\omega_i(k; {\boldsymbol{\theta}}^{(m)}) & = \mathbb{P}(R_{i} = k\mid X_{1:n},R_n=K; {\boldsymbol{\theta}}^{(m)})\\
&=\frac{\mathbb{P}(X_{1:i}, R_{i} = k\mid{\boldsymbol{\theta}}^{(m)}) \mathbb{P}(X_{(i+1):n},R_n=K\mid R_{i} = k; {\boldsymbol{\theta}}^{(m)}) }{\mathbb{P}(X_{1:n},R_n=K)}\cdot%\mid{\boldsymbol{\theta}}^{(m)})
\end{align*}
It has been proved in~\cite{bouaziz2018change} (see Supporting material) that
\begin{align*}
\mathbb{Q}(\boldsymbol{\theta}\mid{\boldsymbol{\theta}}^{(m)}) &=  \sum_{i = 1}^n \sum_{k = 1}^K  \omega_i(k; {\boldsymbol{\theta}}^{(m)})\log \big(e_i(k ; \theta_k)\big). 
\end{align*}
The E-step can therefore be implemented after computation of the $\omega_i$'s. This is achieved by means of a forward-backward algorithm.

%%where $\omega_i(k; {\color{cyan}\theta^{\text{old}}}) = \mathbb{P}(R_{i} = k|{\color{black}X}; \theta^{\text{old}})$ weights. %$e_i(k ; \theta) = \mathbb{P}(X_{i}|R_{i} = k; \theta) \hspace{0.2cm}$ represent likelihood terms and $\omega_i(k; {\color{cyan}\theta^{\text{old}}}) = \mathbb{P}(R_{i} = k|{\color{black}X}; \theta^{\text{old}})$ weights.
%
%
%
%\vspace{0.7cm}
%
%The E step of the EM algorithm is then based on the calculation of the different $\omega_i$: %Straightforward calculations lead to 
%$$\omega_i(k;{\color{red}\theta}) = \mathbb{P}(R_{i} = k|{\color{black}X}; \theta) = \frac{\mathbb{P}(X_{1:i}, R_{i} = k|\theta) \mathbb{P}(X_{i+1:n}|R_{i} = k; \theta) }{\mathbb{P}(X|\theta)}, \text{ with } X_{i:i'} = \{X_{i}, X_{i+1}, \dots, X_{i'}\}.$$ 
%\subsubsection{EM algorithm for estimating a HMM}
%\subsubsection{Forward Backward algorithm for the E-step of the EM algorithm}

%\label{sssec:FB_algo_problemEM}

By setting $F_{i}(k; \boldsymbol{\theta}) = \mathbb{P}(X_{1:i}, R_{i} = k\mid \boldsymbol{\theta})$, for $i=1,\ldots,n$ (the so-called forward quantities) and $B_{i}(k; \boldsymbol{\theta}) = \mathbb{P}(X_{(i+1):n},R_n = K\mid R_{i} = k; \boldsymbol{\theta})$, for $i=1,\ldots,n-1$ (the so-called backward quantities), we then have%\equiv F_{i}(R_i = k; \theta) %\equiv B_{i}(R_i = k ; \theta) 
$$\omega_i(k;\boldsymbol{\theta}) = \frac{ F_{i}(k; \boldsymbol{\theta}) B_{i}(k; \boldsymbol{\theta})}{\mathbb{P}(X_{1:n},R_n=K)}\cdot$$
The forward and backward quantities can be recursively computed as follows, for $i = 2, \dots,n, k=1,\ldots,K$:
\begin{align*}
	F_{1}(k; \boldsymbol{\theta}) &=e_1(1; {\theta_1})\mathds 1_{k=1},\\
	B_n(k)&=\mathds 1_{k=K},\\
	F_{i}(k; \boldsymbol{\theta}) &= \sum_{j=k-1}^k F_{i-1}(j ; \boldsymbol{\theta}) \phi_{i}(j,k; \boldsymbol{\theta})\mathds 1_{j\geq 1},\\ %\text{ for }  i = 2, \dots,n, k=1,\ldots,K,\\% \ k = 1, \dots,K,
	B_{i-1}(k ; \boldsymbol{\theta})  &= \sum_{j=k}^{k+1} \phi_{{\color{black}i}}({\color{black}k,j} ; \boldsymbol{\theta}) B_{i}(j; \boldsymbol{\theta})\mathds 1_{j\leq K},% \text{ for }  i = 2, \dots,n,k=1,\ldots,K,
\end{align*}
where
\begin{align*}
	\phi_{i}(j,k;\boldsymbol{\theta}) &= \mathbb{P}(R_i = k, X_{i}\mid R_{i-1} = j; \boldsymbol{\theta})=e_i(k; {\theta_k})\mathbb{P}(R_i = k\mid R_{i-1} = j).%\equiv \phi_{i}(R_{i-1} = j, R_i = k ; \theta) \\
\end{align*}
%
%The forward and backward quantities can be recursively computed as follows \cite{luong2013fast}:
%\begin{align*}
%F_{1}(k; \boldsymbol{\theta}) &= \begin{cases} e_1(1; {\theta_1}) & \text{ if } k=1\\
%0 & \text{ if } k>1\end{cases}\\
%F_{i}(1; \boldsymbol{\theta}) &=e_1(1; {\theta_1})\mathbb P(R_i=1\mid R_{i-1}=1)F_{i-1}(1; \boldsymbol{\theta}), \text{ for }  i = 2, \dots,n,\\
%F_{i}(k; \boldsymbol{\theta}) &= \sum_{j=k-1}^k F_{i-1}(j ; \boldsymbol{\theta}) \phi_{i}(j,k; \boldsymbol{\theta}), \text{ for }  i = 2, \dots,n, k=2,\ldots,K,\\% \ k = 1, \dots,K,
%B_{n-1}(k; \boldsymbol{\theta})  &= \begin{cases} e_n(K; {\theta_K})\mathbb{P}(R_n = K\mid R_{n-1} = K-1) & \text{ if } k=K-1\\
% e_n(K; {\theta_K})& \text{ if } k=K\\
% 0& \text{ if } k<K-1,\end{cases}\\
% B_{i-1}(K ; \boldsymbol{\theta})  &=e_i(K; {\theta_K}) B_{i}(K; \boldsymbol{\theta}), \text{ for }  i = 2, \dots,n-1,\\
% B_{i-1}(k ; \boldsymbol{\theta})  &= \sum_{j=k}^{k+1} \phi_{{\color{black}i}}({\color{black}k,j} ; \boldsymbol{\theta}) B_{i}(j; \boldsymbol{\theta}), \text{ for }  i = 2, \dots,n-1,k=1,\ldots,K-1,
%\end{align*}
%where 
%\begin{align*}
%\phi_{i}(j,k;\boldsymbol{\theta}) &= \mathbb{P}(R_i = k, X_{i}\mid R_{i-1} = j; \boldsymbol{\theta})=e_i(k; {\theta_k})\mathbb{P}(R_i = k\mid R_{i-1} = j).%\equiv \phi_{i}(R_{i-1} = j, R_i = k ; \theta) \\
%%&= \mathbb{P}(R_i = k, X_{i}|R_{i-1} = j; \theta) \\
%%&= \underbrace{\mathbb{P}(X_{i}|R_i; \theta)}_{e_i(k; \theta)} \underbrace{\mathbb{P}(R_i = k|R_{i-1} = j)}_{\pi(j,k)}.
%\end{align*}
%
In practice, these calculations are done in logarithmic scale in order to avoid underflow problems (see Appendix \ref{sssec:annexeFwBwLogScale} for more details).

\textcolor{black}{
	To summarize, the EM algorithm follows the two steps:}%recapitulate, these forward and backward quantities are incorporated in the EM algorithm as follows: 
\begin{itemize}
	\item \textbf{E:} computation of the weights $\omega_i(k; {\boldsymbol{\theta}}^{(m)}) = \mathbb{P}(R_{i} = k\mid X_{1:n}; {\boldsymbol{\theta}}^{(m)}) \propto F_i(k, {\boldsymbol{\theta}}^{(m)}) B_i(k,{\boldsymbol{\theta}}^{(m)})$ (use of the forward-backward algorithm).\\
	%\vspace{0.15cm}
	\item \textbf{M:} update of the parameter value: ${\boldsymbol{\theta}}^{(m+1)} = \underset{\boldsymbol{\theta}}{\operatorname{argmax}} \ \mathbb{Q}(\boldsymbol{\theta}\mid {\boldsymbol{\theta}}^{(m)})$.
	%\item $\theta^{\text{new}} = \underset{\theta}{\operatorname{argmax}} \ \textcolor{red}{\sum_{i = 1}^n \log e_i(R_i^{\text{*}} ; \theta)}$, where $R_i^{\text{*}}$ is from the forward-backward algorithm?
\end{itemize}

%{\color{red}Indiquer ce que EM maximise reellement et indiquer que ce n'est pas la bonne maximisation qu'on souhaite faire / qui permet de calculer correctement les breakpoints}.

%The main drawback of these algorithms based on the traditional EM is that they are not well suited for calculating breakpoints. Indeed, as presented in section ..., such

As mentioned earlier, it is important to stress that this algorithm maximizes with respect to $\boldsymbol\theta$ the observed likelihood
\begin{align}\label{eq:EM_objective}
\mathbb{P}(X_{1:n} \mid\boldsymbol{\theta})=\sum_{R_{1:n}} \mathbb P(X_{1:n}\mid R_{1:n};\boldsymbol{\theta})\mathbb P(R_{1:n}),
\end{align}
where the sum is taken over all possible segmentations such that $R_n=K$. Taking the logarithm of this quantity clearly gives a different expression than the objective quantity defined in Equation~\eqref{eq:likelihood} and the EM algorithm will not provide a maximizer of $\ell_n(\boldsymbol\theta;n_{1:(K-1)})$. Looking at Equation~\eqref{eq:EM_objective}, we see that the EM algorithm makes a compromise by finding the $\boldsymbol{\theta}$ parameter that maximizes the likelihood over all possible segmentations when $\boldsymbol{\theta}$ is shared in all segmentations.

\subsection{The max-EM algorithm for ordered data in a HMM}\label{sec:maxEM_method}

Instead of averaging over all possible segmentations, the max-EM attempts at finding the best possible segmentation and at maximizing the ${\theta_k}$ parameter in each of these segments. 
%calculating the double-sum $\mathbb{Q}(\theta|\theta^{\text{old}}) = \sum_{i = 1}^n \sum_{k = 1}^K  \log e_i(k; \theta) \omega_i(k; \theta^{\text{old}})$ in the step E of the EM algorithm, we can focus on computing maxima leading to the same result. 
For that purpose, we consider 
%
%{\color{orange}first note that computing $$\underset{R}{\operatorname{max}} \ \underset{\theta}{\operatorname{max}} \log e_i(k ) \omega_i(k; \theta^{\text{old}})$$ corresponds to the naive way used in the ``Brute force'' approach. The alternative we consider here consists in rather computing $$\underset{\theta}{\operatorname{max}}\ \underset{R}{\operatorname{max}} \log e_i(k ; \theta) \omega_i(k; \theta^{\text{old}}).$$ % \cite{celeux1992classification} 
%}
%
the max-forward and max-backward quantities that are given, for all $k \in \{1, \dots,K\}$, by%and with $\theta^{\text{old}}$ %similarly to the forward and backward quantities in the previous section, 
\begin{align*}
	F_{i}^{\text{max}}(k ; \boldsymbol{\theta}) &= \max_{R_1, \dots, R_{(i-1)}} \mathbb{P}(R_{1:(i-1)}, R_i = k,{\color{black}X_{1:i}}\mid \boldsymbol{\theta}), \text{ for } i=1,\ldots,n,\\
	B_{i}^{\text{max}}(k;\boldsymbol{\theta}) &= \max_{R_{i+1}, \dots, R_{n-1}} \mathbb{P}(R_{(i+1):(n-1)}, R_n = K,X_{(i+1):n}\mid R_i = k; \boldsymbol{\theta}), \text{ for } i=1,\ldots,n-1,
\end{align*}
respectively. One should note the similarity with the forward and backward quantities introduced in the previous section where the sum symbol has been replaced by the maximum.
%\vspace{0.2cm}
%\noindent 
Furthermore, the max-forward and max-backward quantities can also be explicitly computed using the recurrence formulas: 
\begin{align*}
	F_{i}^{\text{max}}(k ; \boldsymbol{\theta}) &= \max_{j\in\{k-1,k\}} F_{i-1}^{\text{max}}(j; \boldsymbol{\theta}) \phi_{i}(j,k;\boldsymbol{\theta}), \text{ for }  i = 2, \dots,n, k=2,\ldots,K,\\ 
	B_{i-1}^{\text{max}}(k;\boldsymbol{\theta}) &= \max_{j\in\{k,k+1\}}B_{i}^{\text{max}}(j;\boldsymbol{\theta}) \phi_{i}(k,j),, \text{ for }  i = 2, \dots,n-1,k=1,\ldots,K-1,
\end{align*}
with similar formulas for $F_{1}^{\text{max}}(k ; \boldsymbol{\theta})$, $F_{i}^{\text{max}}(1 ; \boldsymbol{\theta}) $, $B_{n-1}^{\text{max}}(k;\boldsymbol{\theta})$, $B_{i-1}^{\text{max}}(K;\boldsymbol{\theta})$ as in the previous section. 
%<!-- ??? Notation k, j ou R_k and R_j ? bizarre ? -->
%<!-- ??? Aussi, B_i correct ?-->
%\noindent with \textcolor{cyan}{$F_1^{\text{max}}(k ;\boldsymbol{\theta}) = \mathbb{P}(X_1, R_1 = k\mid\boldsymbol{\theta}) = \mathbb{P}(X_1\mid R_1 = k\mid\boldsymbol{\theta}) \mathbb{P}(R_1 = k)$} and with the convention that $B_{n}^{\text{max}}(\cdot) \equiv 1$.
%
Given a current parameter ${\boldsymbol{\theta}}^{(m)}$, the quantities $F_{i}^{\text{max}}(k ; {\boldsymbol{\theta}}^{(m)})$ and $B_{i}^{\text{max}}(k;{\boldsymbol{\theta}}^{(m)})$ are then combined to compute the Maximum a Posteriori (MAP): 
\begin{align}\label{eq:maxFB}
	F_{i}^{\text{max}}(k ; {\boldsymbol{\theta}}^{(m)}) B_{i}^{\text{max}}(k;{\boldsymbol{\theta}}^{(m)}) &%\equiv  \underbrace{F_{i}^{\text{max}}(R_{i} = k;\theta) B_{i}^{\text{max}}(R_{i} = k;\theta)}_{ = \text{MAP in } R_{i}}& \\
%%	%&= {\color{orange}\max_{R_{-j}} \mathbb{P}(R_{j},R_{-j},D)} &\\
%%	%&= \max_{R_{-j}} \prod_{i = 1}^{n} \phi_{i}(R_{i-1},R_{i})&  \\
%%	&={\color{cyan}\underset{R_{1:i-1}}{\operatorname{max}} \mathbb{P}(R_1, \dots, R_{i-1}, R_i = k,X_{1:n}, {\color{red}R_n = K}|Z_{1:n})} &\\
	=\underset{R_{1:(i-1)}, R_{(i+1):(n-1)}}{\operatorname{max}} \mathbb{P}(R_{1:(i-1)}, R_i = k,R_{(i+1):(n-1)},X_{1:n},R_n = K \mid{\boldsymbol{\theta}}^{(m)}),
\end{align}

\noindent and from the MAP, we update the segmentation allocation as: $${R_{i}^{\text{max}}}^{(m+1)} = \underset{k}{\operatorname{argmax}} \;F_{i}^{\text{max}}(k ; {\boldsymbol{\theta}}^{(m)}) B_{i}^{\text{max}}(k;{\boldsymbol{\theta}}^{(m)}).$$ 

\noindent Then, in order to update the value of the parameter $\boldsymbol{\theta}$, we maximize, with respect to $\boldsymbol{\theta}$, the quantity 
\begin{align}\label{eq:theta_update}
\sum_{k = 1}^K \sum_{i = 1}^n \log e_i(k ; \theta_k) \mathds{1}_{{R_{i}^{\text{max}}}^{(m+1)}= k}. 
%
%{\color{orange}\Big(\text{or its pondered version } \sum_{i = 1}^n \log e_i(R_i^{\text{max}} ; \theta) \omega_i(R_i^{\text{max}}; \theta^{\text{old}}) \Big)}
\end{align}
Note that, in the above formula, the maximization can be performed for each $\theta_k$ separately by splitting the log likelihood over each segment. % over $\boldsymbol{\theta}$ amounts to maximizing separately over each $\theta_k$ 
The max-forward and max-backward quantities thus lead to the so-called max-EM algorithm. %based on the quantity
%%
%%$${\color{red}\displaystyle{\mathbb{Q}^{\text{max}}(\theta|\theta^{\text{old}}) = \sum_{{\color{red}?}} \underset{R_{1:n}}{\operatorname{max}} \ \log \mathbb{P}(X_{?},R_{?}|\theta) \ \mathbb{P}(R_{?}|{\color{cyan}X_{?}},\theta^{\text{old}})  = \sum_{i = 1}^n \underset{R_{1:n}}{\operatorname{max}}  \log e_i(k ; \theta) \omega_i(k; \theta^{\text{old}})}.}$$
%%
%$${\color{red}\displaystyle{\mathbb{Q}^{\text{max}}(\theta|\theta^{\text{old}}) = \sum_{i = 1}^n \underset{R_{1:n}}{\operatorname{max}}  \log e_i(k ; \theta) \omega_i(k; \theta^{\text{old}})}.}$$
%%
To summarize, its E- and M-steps proceed as follows:
\begin{itemize}
	\item \textbf{E-step:} 
	\begin{itemize}
		\item Computation of $F_{i}^{\text{max}}(k ; {\boldsymbol{\theta}}^{(m)})$ and  $B_{i}^{\text{max}}(k;{\boldsymbol{\theta}}^{(m)})$, for $i = 1,  \ldots, n$.%respectively max Forward and max Backward quantities
		\item Update of the segmentation allocation
		\begin{align*}
{R_{i}^{\text{max}}}^{(m+1)} = \underset{k}{\operatorname{argmax}} \; F_{i}^{\text{max}}(k ; {\boldsymbol{\theta}}^{(m)}) B_{i}^{\text{max}}(k;{\boldsymbol{\theta}}^{(m)}),\; i=1,\ldots,n.
\end{align*}%using the max Forward - max Backward algorithm
	\end{itemize}	
	\item \textbf{M-step:} update of the parameter value 
	\begin{align*}
{\boldsymbol\theta}^{(m+1)} = \underset{\boldsymbol{\theta}}{\operatorname{argmax}} \; \sum_{k = 1}^K \sum_{i = 1}^n \log e_i(k ; \theta_k) \mathds{1}_{{R_{i}^{\text{max}}}^{(m+1)} = k}.
\end{align*}
\end{itemize}
%~~~~~~~~~~~~~~~~~~~~~~~~~~~~~~~~~~~~~~~~~~~~~~~~~~~~~~~~~~~~~~~~~~~~~~~~~~
Even though underflow issues are less problematic with the max-forward max-backward algorithm, those situations can still arise in practice. 
%Note that the underflow problem in max-Forward max-Backward is not as problematic as in the standard Forward Backward algorithm (which somehow ``averages inaccuracies'' due to its sum-product nature...) but can still exist despite the fact that max-Forward (respectively max-Backward) works with maximum values (which are everytime, at least, greater than the corresponding mean values considered in the standard Forward Backward algorithm). However, the maximum values can, in rare cases, still be too small and cause underflow problems. Therefore, we anticipate such situations by adapting the algorithm by putting it on a logarithmic scale. 
The logarithmic scaling is done in a very similar way as in the previous forward-backward algorithm (see Appendix \ref{sssec:annexeMaxFwMaxBwLogScale} for more details). 

%=========    =========
\subsection{Convergence properties of the max-EM algorithm}\label{sec:theo_res_EM} % : MM algorithm
% Dempster, regularities conditions, etc. 

%Theoretical results in \cite{celeux1992classification}: hold / demonstration in our case. 

In the next proposition we show that each iteration of the max-EM algorithm increases the log-likelihood $\ell_n$ defined in Equation~\eqref{eq:likelihood}. The proof is deferred to the Appendix section and is based on the proof from~\cite{celeux1992classification}. The main difference in our proof comes from the structure of the data where the individuals are ordered and the segment indexes that are assumed to follow a HMM. We equivalently denote $(\boldsymbol\theta^{(m)},\boldsymbol{\mathcal C}^{(m)})$ or $(\boldsymbol\theta^{(m)},{n^{(m)}_{1:(K-1)}})$ the parameters values obtained after the $m^{\text{th}}$ step of the max-EM algorithm, where $\boldsymbol{\mathcal C}^{(m)}=(\mathcal C_1^{(m)},\ldots,\mathcal C_K^{(m)})$ and we recall that $\mathcal C_k^{(m)}=\{n_{k-1}^{(m)}+1,\ldots,n_{k}^{(m)}\} $ represents the set of individuals such that ${R_i^{\text{max}}}^{(m)}=k$. %We similarly define $\boldsymbol{\mathcal C}^{*}$ with $\mathcal C_k^{*}=\{n^*_{k-1}+1,\ldots,n_{k}^*\}$ representing the true set of individuals that belong to the segment index $k$. 

\begin{proposition}\label{prop:cvgce_EM}
The sequence of iterates $(\boldsymbol\theta^{(m)},\boldsymbol{\mathcal C}^{(m)})_{m\geq 1}$ generated using the max-EM algorithm satisfies $\ell_n(\boldsymbol{\theta}^{(m+1)};n_{1:(K-1)}^{(m+1)})\geq \ell_n(\boldsymbol{\theta}^{(m)};n_{1:(K-1)}^{(m)})$. Moreover, if for each set $\{n^*_{k-1}+1,\ldots,n^*_k\}$, $k=1,\ldots,K$, the associated log-likelihood $\sum_{i\in \mathcal C^*_k} \log \big(e_i(k;\theta_k)\big)$ has a unique maximum, then the sequence $(\boldsymbol\theta^{(m)},\boldsymbol{\mathcal C}^{(m)})_{m\geq 1}$ converges towards a stationary parameter.%the true parameters $(\boldsymbol\theta^{*},\boldsymbol{\mathcal C}^{*})$.
\end{proposition}

\subsection{Discussion on the algorithm initialization} 
\label{ssec:initialization}
The max-EM algorithm, like the standard EM algorithm and its variants, is sensitive to parameter initialization due to problems of convergence towards local maxima. When we have no information on the parameters value, it is advised to initialize these algorithms with several different initial values and analyze which initialization best maximizes the likelihood. In our setting, the aim is to define a set of $K'-1$ initialization values for the breakpoints, with $K'\geq K$. Once those values are found, we run our max-EM algorithm for all possible combinations of $K-1$ breakpoints among $K'-1$. For each of these combinations, we can start the max-EM algorithm by maximizing Equation~\eqref{eq:theta_update} and then iterate the max-EM algorithm. Among all initializations, the final result is the one with maximum likelihood value.

One way to determine the set of breakpoints initializations is to randomly select them.  In our experience, this strategy leads to inaccurate results even for simple problems with one or two breakpoints unless the number of breakpoints is very large, which, in turn, is problematic as the computation time drastically increases with the value of $K'$. The challenge is therefore to define efficient methods that provide good results with a small set of breakpoint initialization values. In the following, we propose two methods, the first one is based on the Fused-Lasso (FL) algorithm and the other one is based on Binary Segmentation (BS).

%we propose two different methods, one based on the Fused-Lasso (FL) algorithm and the other one based on Binary segmentation. Each of these methods will 
%
% $n_1,\ldots,n_K$. 
% for 
%We propose two different methods for 
%
%This initialization corresponds, in our context of breakpoint calculation, to consider several different sets of $K-1$ initial breakpoints and to initialize each of the $K$ associated segments using some statistics (for instance, based on mean values) computed from the available data. %moments and / or cumulants
%%
%A common approach is to simply use a large number of initializations with random values. 
%%
%This strategy is not really adapted to our context because it leads to relatively inaccurate results (even for simple problems with one or two breakpoints) with a number of initializations which explodes with the number of breakpoints (it would be necessary to consider several different combinations of initial values for higher number $K$ of segments). 
%%
%As an alternative, we propose the two following approaches.

\subsubsection{Fused-Lasso initialization} 

Our first approach uses the Fused-Lasso (FL) algorithm~\citep[see][]{tibshirani2005sparsity, rinaldo2009properties} in the overparameterized model where the number of segments is equal to the number of individuals. The selection of candidate breakpoints goes through the following steps. 

\begin{enumerate}
\item First, solve the problem
\begin{align*}
\boldsymbol{\theta}\in\argmax_{\theta_1,\ldots,\theta_n}\left\{ \sum_{i=1}^n \log(e_i(i;\theta_i))-\lambda\sum_{j=1}^d\sum_{i=1}^{n-1} |\theta^j_{i+1}-\theta^j_i|\right\},
\end{align*}
where $\lambda>0$ is a penalty term and $\theta^j_i$ represents the $j$th component of the $d$-dimensional $\theta_i$ parameter. This is simply a penalized version of Equation~\eqref{eq:likelihood} where $K=n$ and $\mathcal C_k=k$ for $k=1,\ldots,n$. We implement this FL problem using the \texttt{glmnet} R package by rewriting it in terms of a standard Lasso problem through the parametrization $(\theta^1_1,\ldots,\theta^1_n,\ldots,\theta^d_1\ldots,\theta^d_n)^{\top}=D\gamma$ where $D$ is a block matrix of size $dn \times dn$ whose $d$ diagonal blocks are equal to a lower triangular matrix with nonzero elements equal to $1$ and whose $d^2-d$ off-diagonal blocks are equal to matrices of zeros. See~\cite{bouaziz2015penalized} for an example of such implementation of the FL algorithm.%See 
%\begin{align*}
%\boldsymbol{\gamma}\in\argmax_{\gamma_1,\ldots,\gamma_n} \sum_{i=1}^n \log(e_i(i;(D\gamma)_i))+\lambda\sum_{j=1}^d\sum_{i=2}^n |\gamma^j_i|,
%\end{align*}
%where $\theta=D\gamma$, $\theta=(\theta^1_1,\ldots,\theta^1_n,\ldots,\theta^d_1\ldots,\theta^d_n)^{\top}$ and $D$ is a block matrix of size $dn \times dn$ whose $d$ diagonal blocks are equal to a lower triangular matrix with nonzero elements equal to $1$ and $d^2-d$ off-diagonal blocks equal to matrices of zeros.
%
%This fused-lasso problem can be simply  which allows us to use the \texttt{glmnet} R package. This is done with 
\item With the \texttt{glmnet} R package, the problem is solved for a grid of $\lambda$ values. Each of these values corresponds to a number of different $\theta$ parameters: when all $\theta_i^j$ are different from all $\theta_{i+1}^j$ parameters, we consider that the distribution of the data is different between the two segments. For a high penalty value, all $\theta_i^j$ are equal to all $\theta_{i+1}^j$, for $i=1,\ldots,n-1$ and there is only one segment. As the penalty value decreases, the number of segments increases. 
Based on this regularization path we choose the maximum value of $\lambda$ that corresponds to a number of segments equal to at least $5(K-1)$ breakpoints (that is at least $5(K-1) + 1$ segments).
%The number of segments associated with the different values of $\lambda$ is called the
\item We conclude by removing the breakpoints that are too close to each other. We set a minimum number of individuals per segment equal to $50$ and as long as this criterion is not met, we sequentially remove breakpoints starting from the breakpoints that are the closest to each other. We also impose to keep at least $ \lfloor \frac{3}{2}(K-1) \rfloor$ breakpoints in this final selection.
\end{enumerate}
Once this step is finished, we end up with a set of $K'-1$ potential breakpoints for the initialization of the max-EM algorithm. We will then run our max-EM algorithm for all possible combinations of $K-1$ breakpoints among $K'-1$. This means our algorithm will be run $K'-1 \choose {K-1}$ times. The threshold values $5(K-1)$ and $ \lfloor \frac{3}{2}(K-1) \rfloor$ used in steps $2.$ and $3.$, respectively, are arbitrary and were chosen based on simulation experiments. They seem to provide a good compromise between the need to explore a large number of initializations and computer complexity. In our simulation experiments, this strategy was working with scenarios up to $7$ breakpoints. Of note, the algorithmic complexity for FL is of order $\mathcal O(n^2)$ in our case, since the penalization is applied to $n-1$ consecutive differences. Also, the total computation time is sensitive to the type of regression modeling that is implemented (typically, a Poisson regression model is more computer intensive than a linear model).% say $a$,

\subsubsection{Binary Segmentation initialization} 
Our second approach is based on the Binary Segmentation (BS) strategy~\citep[see][]{scott1974cluster}. The idea is based on a recursive splitting of the data and application of the max-EM algorithm in the one breakpoint situation. We start by running the one breakpoint max-EM algorithm, where the breakpoint is initialized at the middle of the sample. Once this is done we separately consider the two sub-samples made by the two segments and we apply twice the one breakpoint max-EM algorithm in each of those sub-samples. Again, the max-EM is initialized by setting the initial breakpoint as the middle value of the sub-sample. This recursion is applied four times which provides us with a total of $1 + 2 + 4 + 8=15$ breakpoints. As before, we run our max-EM algorithm for all possible combinations of $K-1$ breakpoints among $15$. The number of recursions is arbitrary and is based on simulation experiments. It is important to stress that the one breakpoint max-EM algorithm is extremely fast to run, of order $\mathcal O(n)$, and the whole procedure needed to define our set of breakpoint initializations requires $15$ calls of our one breakpoint model. While it might be possible to reduce this number when $K$ is small, it is rather convenient in practice to simply fix this value. This gives a computational advantage of the BS initialization over FL.

On the other hand, the set of initial breakpoints will tend to be larger with the BS method than with the FL method, which will also impact the computation time of the two strategies.

\subsection{Inferring the number of breakpoints $K$ with BIC}% in max-EM
\label{ssec:bicInferNBbp}
The methodology developed so far works only for a fixed number of $K$. In this section, we propose to use the Bayesian Information Criterion (BIC) to infer this value, as in~\cite{bouaziz2018change}. The criterion has the following form: 
%the right number of breakpoints computed by the max-EM algorithm like in~\cite{bouaziz2018change}. 
$$-2\ell_n(\boldsymbol{\hat{\theta}};n_{1:(K-1)}) + d \times K \times \log (n),$$
where $\boldsymbol{\hat{\theta}}$ is the estimated parameter using our max-EM algorithm and $d \times K$ is the number of estimated parameters. We will choose the value of $K$ that minimizes this criterion. In practice, this means that we will need to run our max-EM algorithm (including the initialization strategy) for a sequence of values for $K$ in order to find the final model and estimated parameters.

\section{Statistical test for the one breakpoint situation}\label{sec:test}

In this section we provide a statistical test for the two hypothesis~\eqref{eq:test} in the one breakpoint scenario. For likelihood based methods, a simple statistical test is the likelihood ratio which is defined in the following way. Let  $\ell_n^{H_0}$ be the log-likelihood under $(H_0)$ and $\ell_n^{H_1}$ be the log-likelihood under $(H_1)$, that is
\begin{align*}
\ell_n^{H_0}&=\sup_{\theta}\tilde\ell_n(\theta)=\sup_{\theta} \left\{\sum_{i=1}^{n} \log\left(\mathbb P(X_i;\theta)\right)\right\},\\
\ell_n^{H_1}&=\max_{n_1}\sup_{\theta_1,\theta_2}  \ell_n(\theta_1,\theta_2;n_1)=\max_{n_1}\sup_{\theta_1,\theta_2} \left\{\sum_{i=1}^{n_1} \log\left(\mathbb P(X_i;\theta_1)\right)+ \sum_{i=n_1+1}^{n} \log\left(\mathbb P(X_i;\theta_2)\right)\right\},%\cdot
\end{align*} 
where we have introduced the notation $\tilde\ell_n$ to represent the likelihood in the no-breakpoint model. Note also that, for the sake of simplicity, the $R_i$ term was dropped in the notation $\mathbb P(X_i;\theta_k)$ to denote the probability distribution function $\mathbb P(X_i\mid R_i=k;\theta_k)$, $k=1, 2$. We also recall that our methodology works for discrete or continuous random variables. In the above equations, the supremum is taken over $\theta, \theta_1, \theta_2 \in\Theta$ and the maximum is taken over $n_1\in\{1,\ldots,n-1\}$. The test statistic is then defined as $T_n=2(\ell_n^{H_1}-\ell_n^{H_0})$. 

For a fixed value of $n_1$, we define $(\hat\theta_1,\hat\theta_2)=\argmax_{\theta_1,\theta_2}\ell_n(\theta_1,\theta_2;n_1)$ and $\hat\theta_0=\argmax_{\theta}\tilde\ell_n(\theta)$. It should be noted that $\hat\theta_1$ and $\hat\theta_2$ depend on the value of $n_1$, even though this does not appear in the notation for the sake of simplicity. The test statistic can then be rewritten as $T_n=\max_{n_1}\{2(\ell_n(\hat\theta_1,\hat\theta_2;n_1)-\tilde\ell_n(\hat\theta_0))\}$. In Theorem~\ref{th:likeli_test}, the asymptotic distribution of $2(\ell_n(\hat\theta_1,\hat\theta_2;n_1)-\tilde\ell_n(\hat\theta_0))$ is provided under $(H_0)$, when assuming that $n_1$ and $n-n_1$ converge towards infinity. In the following, we define the estimator of the Fisher information under $(H_0)$:
\begin{align*}
\hat I(\hat\theta_0)=-\frac 1n\sum_{i=1}^{n}\nabla^2 \log(\mathbb P(X_i;\hat{\theta}_0)),
\end{align*}
and we use the notation $u^{\otimes 2}=u^{\top}u$.

%When $n_1$ and $n-n_1$We derive the asymptotic distribution of $T_n$ under $(H_0)$ in the next theorem.

\begin{theorem}\label{th:likeli_test}
Let $n, n_1\in\mathbb N^*$, such that $n>n_1$ and $n_1\to\infty$, $n-n_1\to\infty$. Then, under standard assumptions for maximum likelihood theory, 
\begin{align*}
& 2(\ell_n(\hat\theta_1,\hat\theta_2;n_1)-\tilde\ell_n(\hat\theta_0))\\
&\quad = \frac{n-n_1}{nn_1}  \left[\left(\hat I(\hat\theta_0)\right)^{-1/2}\sum_{i=1}^{n_1}\nabla \log\left(\mathbb P(X_i;\hat{\theta}_0)\right)\right]^{\otimes 2}\\
&\qquad +\frac{n_1}{n(n-n_1)}  \left[\left(\hat I(\hat\theta_0)\right)^{-1/2}\sum_{i=n_1+1}^{n}\nabla \log\left(\mathbb P(X_i;\hat{\theta}_0)\right)\right]^{\otimes 2}\\
&\qquad -\frac{2}{n} \left(\sum_{i=1}^{n_1}\nabla \log\left(\mathbb P(X_i;\hat{\theta}_0)\right)\right)^{\top} \left(\hat I(\hat\theta_0)\right)^{-1}\left(\sum_{i=n_1+1}^{n}\nabla \log\left(\mathbb P(X_i;\hat{\theta}_0)\right)\right)+o_{\mathbb P}(1).
\end{align*}
%%Assume that for $j=1,\ldots,J$, there exists $0<\lambda_j<1$ such that $n_j/n$ tends to $\lambda_j$. Then 
%the likelihood ratio test statistic $T_n$ converges in distribution towards%=2(\ell_n^{H_1}-\ell_n^{H_0})
%\begin{align*}
%\max_{\lambda_j} \left\{Z_1^{\top}Z_1 (1-\lambda_j)+Z_2^{\top}Z_2 \lambda_j-2Z_1^{\top}Z_2\sqrt{\lambda_j (1-\lambda_j)}\right\},
%\end{align*}
%where $Z_1$, $Z_2$ are two independent random gaussian vectors of dimension $d$ with mean $0_d$ and variance equal to identity.
\end{theorem}

This theorem can be used to compute the distribution of $T_n$ under $(H_0)$ when $n_1$ and $n-n_1$ are large. The approximation of $2(\ell_n(\hat\theta_1,\hat\theta_2;n_1)-\tilde\ell_n(\hat\theta_0))$ provided by the theorem can be computed for a sequence of $n_1$ values in an efficient way, then taking the maximum over this sequence will provide an approximation of $T_n$. It should be noted that the approximation does not depend on the estimators $\hat\theta_1$ and $\hat\theta_2$; only estimators in the no-breakpoint model must be computed. In practice, the estimator $\hat\theta_0$, the estimator of the Hessian matrix based on the whole sample and evaluated at $\hat\theta_0$, the estimator of the score vector $\nabla \log(\mathbb P(X_i;\hat{\theta}_0))$ for $i=1,\ldots,n$ are fast to compute.%for all subsamples $1,\ldots,n_1$ and $n_1+1,\ldots,n$%ese quantities%As a matter of fact, i%is achieved in a fast way.

When $n_1$ or $n-n_1$ are small, the remainder term in the approximation will no longer be small and this approximation should not be used. The next theorem provides two new approximations for $2(\ell_n(\hat\theta_1,\hat\theta_2;n_1)-\tilde\ell_n(\hat\theta_0))$ corresponding to these two settings.

\begin{theorem}\label{th:likeli_test_small}
Let $n, n_1\in\mathbb N^*$, such that $n>n_1$.
\begin{enumerate}
\item Under standard assumptions for maximum likelihood theory, if $n_1$ is fixed and $n\to\infty$ then
\begin{align*}
2(\ell_n(\hat\theta_1,\hat\theta_2;n_1)-\tilde\ell_n(\hat\theta_0))& = 2\sum_{i=1}^{n_1}\left\{\log\left(\mathbb P(X_i;\hat{\theta}_1)\right)-\log\left(\mathbb P(X_i;\hat{\theta}_0)\right)\right\}+o_{\mathbb P}(1).
%&\quad -\frac{1}{n-n_1}  \left[\left(\hat I(\hat\theta_0)\right)^{-1/2}\sum_{i=n_1+1}^{n}\nabla \log\left(\mathbb P(X_i;\hat{\theta}_0)\right)\right]^{\otimes 2}\\
%&\quad+o_{\mathbb P}(1).
%%&\qquad -\frac{1}{n}  \left[\left(\hat I(\hat\theta_0)\right)^{-1/2}\sum_{i=1}^{n_1}\nabla \log\left(\mathbb P(X_i;\hat{\theta}_0)\right)\right]^{\otimes 2}\\
%%&\qquad -\frac{n_1}{n(n-n_1)}  \left[\left(\hat I(\hat\theta_0)\right)^{-1/2}\sum_{i=n_1+1}^{n}\nabla \log\left(\mathbb P(X_i;\hat{\theta}_0)\right)\right]^{\otimes 2}\\
%&\qquad -\frac{2}{n} \left(\sum_{i=1}^{n_1}\nabla \log\left(\mathbb P(X_i;\hat{\theta}_0)\right)\right)^{\top} \left(\hat I(\hat\theta_0)\right)^{-1}\left(\sum_{i=n_1+1}^{n}\nabla \log\left(\mathbb P(X_i;\hat{\theta}_0)\right)\right)+o_{\mathbb P}(1).
\end{align*}
\item Under standard assumptions for maximum likelihood theory, if $n_1\to\infty$ and $n-n_1$ converges towards a positive constant, then
\end{enumerate}
\begin{align*}
2(\ell_n(\hat\theta_1,\hat\theta_2;n_1)-\tilde\ell_n(\hat\theta_0))&= 2\sum_{i=n_1+1}^{n}\left\{\log\left(\mathbb P(X_i;\hat{\theta}_2)\right)-\log\left(\mathbb P(X_i;\hat{\theta}_0)\right)\right\}+o_{\mathbb P}(1).
%&\quad -\frac{1}{n_1}  \left[\left(\hat I(\hat\theta_0)\right)^{-1/2}\sum_{i=1}^{n_1}\nabla \log\left(\mathbb P(X_i;\hat{\theta}_0)\right)\right]^{\otimes 2}\\
%&\quad+o_{\mathbb P}(1).
%&\qquad +\frac{n_1}{n(n-n_1)}  \left[\left(\hat I(\hat\theta_0)\right)^{-1/2}\sum_{i=n_1+1}^{n}\nabla \log\left(\mathbb P(X_i;\hat{\theta}_0)\right)\right]^{\otimes 2}\\
%&\qquad +\frac{2}{n} \left(\sum_{i=1}^{n_1}\nabla \log\left(\mathbb P(X_i;\hat{\theta}_0)\right)\right)^{\top} \left(\hat I(\hat\theta_0)\right)^{-1}\left(\sum_{i=n_1+1}^{n}\nabla \log\left(\mathbb P(X_i;\hat{\theta}_0)\right)\right)+o_{\mathbb P}(1).
\end{align*}
\end{theorem}
As opposed to Theorem~\ref{th:likeli_test}, those results require the computation of $\hat\theta_1$ and $\hat\theta_2$. However, the idea is to use 1. of Theorem~\ref{th:likeli_test_small} for small values of $n_1$ (typically less than $100$) and to use 2. of Theorem~\ref{th:likeli_test_small} for small values of $n-n_1$ (typically less than $100$). We will therefore combine Theorems~\ref{th:likeli_test} and~\ref{th:likeli_test_small} to compute $\{2(\ell_n(\hat\theta_1,\hat\theta_2;n_1)-\tilde\ell_n(\hat\theta_0))\}$ for all values of $n_1$ and take the maximum to derive $T_n$. The proofs of those two theorems are provided in the Appendix section.%This strategy will not work well if the estimator properties of the regression parameters are only asymptotic. %to check

\section{Simulations}\label{sec:simu}

In the following, we will evaluate the performance of our method in various simulation settings. In Section~\ref{ssec:bpDetectionSimu1} we consider a simple mean model which allows comparisons of our method with the Brute-Force method (in the one breakpoint situation) and the GFPOP algorithm. In Section~\ref{ssec:bpDetectionSimu2}, three regression models are considered: a linear, a logistic and a survival models. In Section~\ref{ssec:bpDetectionSimu3}, the power of the statistical test developed in Section~\ref{sec:test} is investigated in the three previous regression models with one breakpoint.

All the simulations are replicated on $J=500$ samples. For $j=1,\ldots,J$, $k=1,\ldots, K$, let $\hat\theta_k^{(j)}=(\hat\theta_{k,1}^{(j)},\ldots,\hat\theta_{k,d}^{(j)})^{\top}\in\mathbb R^d$ denote the estimate of the true parameter $\theta_k^*=(\theta_{k,1}^*,\ldots,\theta_{k,d}^*)^{\top}$ in segment $k$, obtained from the $j$th Monte Carlo sample. In order to assess the performance of this estimator, the Mean Squared Error (MSE) decomposed as the sum of the variance (VAR) and the squared bias $\text {BIAS}^2$, and the Mean Absolute Percentage Error (MAPE) are used as metrics. They are defined in the following way:%=(\hat\theta^{(m),1},\ldots,\hat\theta^{(m),d})
\begin{align*}
\text {MSE}( \boldsymbol{\hat{\theta}})&=\frac 1{K J} \sum_{j=1}^J \sum_{k=1}^K(\hat\theta_k^{(j)}-\theta_k^*)^{\top}(\hat\theta_k^{(j)}-\theta_k^*)\\
\text {BIAS}^2( \boldsymbol{\hat{\theta}})&=\frac 1{K}\sum_{k=1}^K(\bar{\hat\theta}_k-\theta_k^*)^{\top}(\bar{\hat\theta}_k-\theta_k^*)\\
\text {VAR}( \boldsymbol{\hat{\theta}})&=\frac 1{KJ} \sum_{j=1}^J\sum_{k=1}^K (\hat\theta_k^{(j)}-\bar{\hat\theta}_k)^{\top}(\hat\theta_k^{(j)}-\bar{\hat\theta}_k)\\
\text {MAPE}( \boldsymbol{\hat{\theta}})&=\frac 1{K J} \sum_{j=1}^J \sum_{k=1}^K\sum_{l=1}^d \left|\frac{\hat\theta_{k,l}^{(j)}-\theta_{k,l}^*}{\theta_{k,l}^*}\right|,%d 
\end{align*}
where $\bar{\hat\theta}_k=\sum_j \hat\theta_k^{(j)}/J$. Contrary to the MSE, bias and variance, the MAPE metric takes into account the amplitude of the parameter values. On the other hand, the accuracy error of breakpoints detection is evaluated through the criterion:%All criterions are normalised by the number of estimated parameters $d$. 
\begin{align*}
\text {ACCE}(\text{bp})=\frac 1{n} \sum_{j=1}^J\sum_{i=1}^n \mathds 1_{\hat R^{(j)}_i\neq R_i},%\sum_{j=1}^J \sum_{k=1}^K \left|\hat{\text{bp}}_k^{(j)}-\text{bp}_k^*\right|,
\end{align*}
where $\hat R^{(j)}_i$ is the estimated segment index for individual $i$ in sample $j$ and we recall that $R_i$ is the true segment index for individual $i$. Therefore, this metric evaluates the proportion of individuals that are allocated the incorrect segment index.
%The value MAPE(bp) is also computed for the Mean Absolute Percentage Error for the breakpoint values using the same formula where we average over all replications and all breakpoints.

%=========    =========
\subsection{Implementation of the max-EM algorithm in the mean model}%Breakpoint detection in simulated data for different {\color{red}loss functions}
\label{ssec:bpDetectionSimu1}

In this section we consider the simple following model:
\begin{align*}
\text{for } k = 1, \ldots,K, \  i = n_{k-1}^*+1, \dots, n_{k}^*, \ Y_i = \theta_k^{*} + \varepsilon_i,
\end{align*}
%\begin{align*}
%Y_i = \theta_k^{*} + \varepsilon_i, \  i = n_{k-1}^*+1, \dots, n_{k}^* \text{ and } k = 1, \ldots,K,
%\end{align*}
where $\varepsilon_i \sim \mathcal{N}(0, \sigma^{*2})$ and $\theta_k^{*}\in\mathbb R$. This is an homoscedastic model since the variance $\sigma^{*2}$ is assumed to be equal for all $K$ segments. The aim of this simulation setting is first, to compare the two proposed initialisations, the one based on the Fused Lasso (FS) and the other based on Binary Segmentation (BS) and second, to compare our implementations with the Brute Force method and with the GFPOP algorithm. For this second goal, the comparison with brute force can only be made in a one breakpoint situation (that is when $K=2$) due to computational issues arising for $K\geq 3$. We consider two settings, one with one breakpoint ($K=2$) and another setting with $5$ breakpoints ($K=6$). %In both settings the sample size is equal to $n=1000$.
\begin{itemize}
\item One breakpoint: $\theta_1^{*}=10$, $\theta_2^{*}=12$, $\sigma^{*}=3$, $n=500$ and $n_1^*=345$.
\item Five breakpoints: $\theta_1^{*}=19$, $\theta_2^{*}=23$,  $\theta_3^{*}=30$, $\theta_4^{*}=35$, $\theta_5^{*}=42$, $\theta_6^{*}=37$, $\sigma^{*}=5$, $n=1,000$ and $n_1^*=82$, $n_2^*=333$, $n_3^*=508$, $n_4^*=701$, $n_5^*=945$. 
\end{itemize}
The results are presented in Table~\ref{tab:homo_one_five_bp} where the MSE of the algorithms are provided along with its decomposition as the sum of the variance and the squared bias. The MAPE of the parameters and of the breakpoints values is also computed. 

\begin{figure}[htbp]
	\begin{center}
		\includegraphics[height=0.5\linewidth]{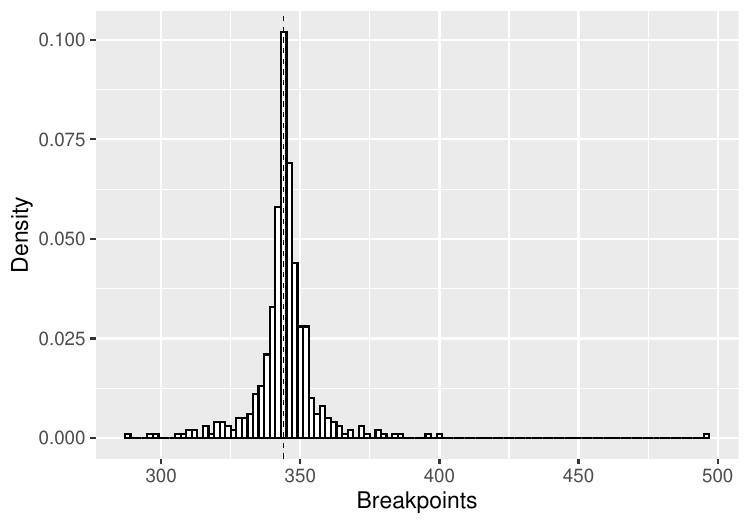}
	\end{center}
\caption{\small{Distribution of breakpoints computed in the one breakpoint homoscedastic mean model. The distribution was obtained based on $500$ replications and is identical for the max-EM with BS initialization, GFPOP and Brute Force algorithms. The vertical dotted line indicates the true breakpoint, equal to $345$ in this simulation setting.}}   
	\label{fig:gauss_homosced_1_bp_withoutCov_BS_histogram}
\end{figure}

In the one breakpoint setting we first observe that all three methods (max-EM with BS initialization, GFPOP and Brute Force) have the same performance for the proposed metrics. In fact, the estimates for all $J=500$ samples are identical. On the other hand, the max-EM with FL initialization provides very similar results: indeed, by looking more closely at the results, it turns out that, out of the $500$ replications, there is only one sample where max-EM with FL initialization provides a different breakpoint than the other methods. For this breakpoint, it finds the breakpoint $\hat{n}_1= 319$ with corresponding likelihood-value equal to $-2,498.02$, when all the other methods find the breakpoint $\hat{n}_1= 343$  with corresponding likelihood-value equal to $-2,497.98$ (we recall that the true breakpoint is $n_{1}^* = 345$). The distribution of the estimated breakpoint based on all three methods is also provided in Figure~\ref{fig:gauss_homosced_1_bp_withoutCov_BS_histogram}. It shows that the algorithms are extremely accurate in terms of breakpoint detection in this setting. Finally, the MSE for the standard deviation of the residuals is equal to $0.00442$ for both max-EM algorithms and for the Brute Force method. We have also compared the computation time of the whole method based on the two initializations, with a clear advantage of the max-EM with BS initialization which runs on average in 2.5 seconds over max-EM with FL initialization which runs on average in 4.5 seconds.

In the five breakpoint setting, all methods provide a very accurate estimation of the parameters based on all metrics. However, the max-EM algorithm with FL initialization tends to be less performant: its variance is twice as big as the variance of the other methods. This highlights the fact that this method sometimes find a sequence of breakpoints that are far from the truth, a phenomenon that does not occur with max-EM with BS initialization and GFPOP whose performances are very similar according to all metrics. Finally, the MSE for the standard deviation of the residuals is equal to $0.00139$ for the max-EM algorithm with FL initialization and to $0.00094$ for the max-EM algorithm with BS initialization.

In light of these results, our algorithm max-EM with BS initialization seems to provide the best tradeoff between accuracy and speed, since its computational cost is linear. In the next simulations, we will only present the results for the BS initialization in the main text, the results for the FL initialization can be found in Supplementary Material.% nevertheless, 

%\begin{table}[ht]
%\centering
%\resizebox{\textwidth}{!}{
%\begin{tabular}{|r|cc|ccc|}
%  \hline
%&\multicolumn{2}{c|}{One bp} & \multicolumn{3}{c|}{Five bp}\\
%%\hline 
%  &max-EM(FL)&max-EM(BS)/GFPOP/BF& max-EM(FL)& max-EM(BS) & GFPOP \\
% \hline
%$ \text {MSE}( \boldsymbol{\hat{\theta}})$ & 0.01834&0.01837 & 0.61054 &0.27507 & 0.23313 \\ 
%  $ \text {BIAS}^2( \boldsymbol{\hat{\theta}})$ & 0.00006&0.00006 & 0.00417 &0.00265 & 0.00463 \\ 
%  $ \text {VAR}( \boldsymbol{\hat{\theta}})$ & 0.01828&0.01831 & 0.60637 &0.27241 & 0.22850 \\ 
%  $ \text {MAPE}( \boldsymbol{\hat{\theta}})$ & 0.00999&0.00999 & 0.01467&0.01318 & 0.01298 \\ 
%  MAPE(bp) & 0.01972&0.01958 & 0.02484&0.01637 & 0.01671 \\ 
%   \hline
%\end{tabular}}
%\caption{\small Results in the simple homoscedastic mean model with two scenarios: the one and five breakpoint models. The Mean Squared Error (MSE) of the estimated mean parameters, decomposed as the variance (VAR) plus squared bias ($\text{BIAS}^2$), 
%                       along with the MAPE of the estimated parameters and of the estimated breakpoints are provided. The max-EM algorithms with Fused Lasso (FL) and Binary Segmentation (BS) initialisations are compared 
%                       with the GFPOP  and Brute Force algorithms.}%The data where simulated following a linear regression model with homoscedastic error, five breakpoints and one mean parameter in each segment 
%\label{tab:homo_one_five_bp}
%\end{table}

\begin{table}[ht]
\centering
\resizebox{\textwidth}{!}{
\begin{tabular}{|r|cc|ccc|}
  \hline
&\multicolumn{2}{c|}{One bp} & \multicolumn{3}{c|}{Five bp}\\
%\hline 
  &max-EM(FL)&max-EM(BS)/GFPOP/BF& max-EM(FL)& max-EM(BS) & GFPOP \\
 \hline
$ \text {MSE}( \boldsymbol{\hat{\theta}})$ & 0.03668&0.03674 & 3.66324 &1.65040 & 1.39877 \\ 
  $ \text {BIAS}^2( \boldsymbol{\hat{\theta}})$ & 0.00012&0.00012 & 0.02502 &0.01592 & 0.02778 \\ 
  $ \text {VAR}( \boldsymbol{\hat{\theta}})$ & 0.03656&0.03662 & 3.63822 &1.63449 & 1.37099 \\ 
  $ \text {MAPE}( \boldsymbol{\hat{\theta}})$ & 0.01997&0.01998 & 0.08801&0.07906 & 0.07787 \\ 
  ACCE(bp) & 0.00680&0.00675 & 0.02756&0.01567 & 0.01449 \\ 
   \hline
\end{tabular}}
\caption{\small Results in the simple homoscedastic mean model with two scenarios: the one and five breakpoint models. The Mean Squared Error (MSE) of the estimated mean parameters, decomposed as the variance (VAR) plus squared bias ($\text{BIAS}^2$) along with the MAPE of the estimated parameters and the ACCE of the estimated breakpoints are provided. The max-EM algorithm is compared with the GFPOP  and brute force algorithms.}%The data where simulated following a linear regression model with homoscedastic error, five breakpoints and one mean parameter in each segment 
\label{tab:homo_one_five_bp}
\end{table}

\subsection{Implementation of the max-EM algorithm in regression models}%Breakpoint detection in simulated data for different {\color{red}loss functions}
\label{ssec:bpDetectionSimu2}

In this section we consider three different regression models in different settings. A linear, a logistic and a survival regression models are studied based on scenarios with one and four breakpoints and several covariates. The models are described in details below.
\begin{itemize}
\item Model 1. Linear regression.
\begin{align*}
\text{For } k = 1, \ldots,K,i = n_{k-1}^*+1, \dots, n_{k}^*, \  Y_i = X_i^{\top}\theta_k^{*} + \varepsilon_i,% \  i = n_{k-1}^*+1, \dots, n_{k}^* \text{ and } k = 1, \ldots,K,
\end{align*}
where $X_i=(1,X_{i,1},X_{i,2})^{\top}$, $X_{i,1}$, $X_{i,2}$ are independent and follow a uniform distribution on $[0,1]$ and $\varepsilon_i$ follows a centered normal distribution with variance $\sigma^2$ equal to $6.25$.% $\theta_k^{*}=()$, $\varepsilon_i$ follows  
\item Model 2. Logistic regression.
\begin{align*}
\text{For } k = 1, \ldots,K,i = n_{k-1}^*+1, \dots, n_{k}^*, \  \mathbb P[Y_i=1\mid X_i] = \frac{\exp(X_i^{\top}\theta_k^{*})}{1+\exp(X_i^{\top}\theta_k^{*})}, %\  i = n_{k-1}^*+1, \dots, n_{k}^* \text{ and } k = 1, \ldots,K,
\end{align*}
where $X_i=(1,X_{i,1})^{\top}$ and $X_{i,1}$ follows a Bernoulli distribution with parameter $p=0.5$. 
\item Model 3. Accelerated Failure Time/Cox regression.
\begin{align*}
\text{For } k = 1, \ldots,K,i = n_{k-1}^*+1, \dots, n_{k}^*, \  \log(Y_i)= X_i^{\top}\theta_k^{*} + \sigma\varepsilon_i,% \  i = n_{k-1}^*+1, \dots, n_{k}^* \text{ and } k = 1, \ldots,K,
\end{align*}
where $X_i=(1,X_{i,1},X_{i,2})^{\top}$, $X_{i,1}$, $X_{i,2}$ are independent and follow a uniform distribution on $[0,1]$, $\varepsilon_i$ has a probability density function equal to $f_{\varepsilon}(w)=\exp(w-\exp(w))$ and $\sigma\in\mathbb R$ is an extra scale parameter. In this model, the outcome $Y_i$ is not directly observed but instead we observe the variable $T_i=Y_i\wedge C_i$, with $C_i$ a censoring variable following an exponential distribution with parameter equal to $0.1$ (that is with expectation equal to $10$). With this censoring distribution, $35\%$ of observations are censored on average. It is important to stress that even though this model is presented as an accelerated failure time model, it can also be recast into a Cox proportional hazard model~\citep[see][]{kalbfleisch2011statistical}. Let $\lambda(\cdot\mid X_i)$ be the conditional hazard rate for the variable $Y_i$, then Model 3 is equivalent to assuming:
\begin{align*}
\text{for } k = 1, \ldots,K, \  i = n_{k-1}^*+1, \dots, n_{k}^*, \ \lambda(t\mid X_i) = \lambda_0(t) \exp(\tilde X_i^{\top}\beta_k^{*}),
\end{align*}
where 
\begin{align*}
\lambda_0(t)=\frac 1{\sigma}\exp\left(-\frac{\theta_{k,1}^{*}}{\sigma}\right)t^{1/\sigma-1},
\end{align*}
$\tilde X_i=(X_{i,1},X_{i,2})^{\top}$ and $\beta_k^{*}=-(1/\sigma)(\theta_{k,2}^{*},\theta_{k,3}^{*})^{\top}$.
\end{itemize}

For each model, a one breakpoint ($K=2$) and two breakpoint ($K=3$) settings are considered. In the one breakpoint setting, all samples are of size $1,000$ and the breakpoints are equal to $553$, $112$ and $666$ in the linear, logistic and survival models, respectively. In the two breakpoint setting, all samples are of size $1,000$ and the breakpoints are equal to $333$ and $666$ in the linear and logistic models, and to $375$ and $689$ in the survival model. The exact values of the parameters in each model and each breakpoint setting are provided in Table~\ref{tab:param_setting_regression}. The results from the max-EM algorithm with BS initialization are presented in Table~\ref{tab:reg_models}. Some of the results with FL initialization can also be found in Supplementary Material. No competitors were computed in those simulation settings: the GFPOP algorithm cannot work with regression models and we were not able to implement the Brute Force algorithm due to computational issues. We observe a good performance of our method in all settings. In particular, the accuracy error of breakpoints detection, ACCE(bp), is extremely low in all settings, which implies that almost all individuals are assigned to the correct segment (the worst situation occurs for the logistic model with two breakpoints in which case ACCE(bp) equals $1.8\%$). Since the max-EM algorithm operates in two steps, with the segment allocation as the first step and separate parameters estimation in each segment as the second step, the parameters estimation error is mainly due to the performance of the maximum likelihood estimators inherent to each model and to the sample size in each segment. In the two breakpoint case, our estimator slightly deteriorates in terms of MSE except for the logistic model. This is due to the balanced setting in terms of number of observations in each segment for the linear and survival models, while for the logistic model, the one breakpoint case is particularly unbalanced with few observations in the first segment ($112$ observations in the first segment and $888$ observations in the second segment). By comparison, in the two breakpoint scenario, there are more observations in all three segments ($333$ in the first two segments and $334$ in the third). Surprisingly, the survival model, that suffers from censoring and has the largest number of parameters, displays the best performance in terms of MSE and breakpoint detection, both in the one breakpoint and two breakpoint settings. In Table 1 of Supplementary Information, we observe that the FL initialization provides slightly better results than BS initialization for the linear and survival models, while for the logistic regression, BS initialization outperforms FL initialization except in terms of bias. In the two breakpoint situation, with the survival model, BS initialization has a slight advantage with all metrics except in terms of bias which is similar for the two initialization methods. %(and is equal to the standard estimation errors obtained in likelihood regression models and is due to the sample size in each segment
 %very similar result as compared to BS initialisation in the one breakpoint situation. However, in the two breakpoint situation, BS initialisation has a slight advantage. 
Considering the computational advantage of BS initialization, those results are in favour of the BS initialization especially when the number of breakpoints is greater than one.

\begin{table}[ht]
\centering
%\resizebox{\textwidth}{!}{
\begin{tabular}{|lr|cc|ccc|}
  \hline
&&\multicolumn{2}{c|}{One bp} & \multicolumn{3}{c|}{Two bp}\\
%\hline 
& & $\theta_1^*$& $\theta_2^*$& $\theta_1^*$& $\theta_2^*$ & $\theta_3^*$  \\
 \hline
Linear &Intercept&1.00 & 2.00 &1.00 & 1.50 &2.00  \\ 
($\sigma=2.5)$ & cov. effect 1&11.40 & 12.30 &11.40 & 5.00 &12.30 \\
 & cov. effect 2&0.60 & 0.10 &0.60 & -1.00 &0.10 \\
Logistic &Intercept&-1.10 & 0.50 &-1.10 & 0.50 &-1.00  \\ 
 & cov. effect &0.60 & -0.20 &0.60 & -0.20 &0.40  \\
 Survival &Intercept&2.00 & 2.50 &2.00 & 2.20 &2.50  \\ 
  & scale &1.70 & 1.98 &1.70 & 1.80 &1.98  \\ 
 & cov. effect 1&3.00 & 3.90&3.00 & 3.40 &3.90  \\ 
  & cov. effect 2&4.20 & 4.90&4.20 & 4.70 &4.90  \\ 
   \hline
\end{tabular}%}
\caption{\small True parameter values in the regression simulation scenarios. A linear, logistic and survival models are studied in the one and two breakpoints settings. The linear model is homoscedastic with an error standard deviation equal to $2.5$ in the two settings. In the one breakpoint setting, the breakpoints are equal to $553$, $112$ and $666$ in the linear, logistic and survival models, respectively. In the two breakpoint setting, the breakpoints are equal to $333$ and $666$ in the linear and logistic models, they are equal to $375$ and $689$ in the survival model.}%The data where simulated following a linear regression model with homoscedastic error, five breakpoints and one mean parameter in each segment 
\label{tab:param_setting_regression}
\end{table}

%\begin{table}[ht]
%\centering
%%\resizebox{\textwidth}{!}{
%\begin{tabular}{|lr|cc|ccccc|}
%  \hline
%&&\multicolumn{2}{c|}{One bp} & \multicolumn{5}{c|}{Two bp}\\
%%\hline 
%& & $\theta_1^*$& $\theta_2^*$& $\theta_1^*$& $\theta_2^*$ & $\theta_3^*$ & $\theta_4^*$ & $\theta_5^*$ \\
% \hline
%Linear &Intercept&1.00 & 2.00 &val1 & val2 &val3 & val4 &val5 \\ 
% & cov. effect 1&11.40 & 12.30 &val1 & val2 &val3 & val4 &val5 \\
% & cov. effect 2&0.60 & 0.10 &val1 & val2 &val3 & val4 &val5 \\
%Logistic &Intercept&-1.10 & 0.50 &val1 & val2 &val3 & val4 &val5 \\ 
% & cov. effect &0.60 & -0.20 &val1 & val2 &val3 & val4 &val5 \\
% Survival &Intercept&2.00 & 2.50 &val1 & val2 &val3 & val4 &val5 \\ 
%  & scale &1.70 & 1.98 &val1 & val2 &val3 & val4 &val5 \\ 
% & cov. effect 1&3.00 & 3.90&val1 & val2 &val3 & val4 &val5 \\ 
%  & cov. effect 2&4.20 & 4.90&val1 & val2 &val3 & val4 &val5 \\ 
%   \hline
%\end{tabular}%}
%\caption{\small Parameter values in the regression simulation scenarios. A Linear, logistic and survival models are studied in the one and four breakpoints settings. The linear model is homoscedastic with an error variance equal to $6.25$ in the two settings. In the one breakpoint setting, with a sample size of $1,000$, the breakpoints are equal to $553$, $112$ and $666$ in the linear, logistic and survival models, respectively.}%The data where simulated following a linear regression model with homoscedastic error, five breakpoints and one mean parameter in each segment 
%\label{tab:param_setting_regression}
%\end{table}

\begin{table}[ht]
\centering
\begin{tabular}{|rr|c|c|c|}
  \hline
 \multicolumn{2}{|c|}{$n=1,000$}  & Linear Model& Logistic Model& Survival Model  \\%\multirow{2}{4.5em}{$n=1,000$}
  && bp\;$=553$& bp\;$=112$& bp\;$=666$  \\
 \hline
\multirow{5}{4.5em}{One bp} & $ \text {MSE}( \boldsymbol{\hat{\theta}})$ & 0.86471 & 1.41481 & 0.10759  \\ %211110.5577
  &$ \text {BIAS}^2( \boldsymbol{\hat{\theta}})$ & 0.00280 & 0.01496 & 0.00157 \\ %335.9970
  &$ \text {VAR}( \boldsymbol{\hat{\theta}})$ & 0.86191 & 1.39566 & 0.10602 \\ %210774.5607
  &$ \text {MAPE}( \boldsymbol{\hat{\theta}})$ & 4.37378 & 2.74543 & 0.26998  \\ %108.9260
  %&MAPE(bp) & 0.02473 & 0.09023 & 0.00240 \\ %0.5023
  &ACCE(bp) & 0.01367 & 0.01011 & 0.00160  \\
   \hline
   \hline
   && bp\;$=(333,666)$& bp\;$=(333,666)$& bp\;$=(375,689)$  \\
     \hline
   \multirow{5}{4.5em}{Two bp} & $ \text {MSE}( \boldsymbol{\hat{\theta}})$ & 1.74872 & 1.27661 &  0.26253  \\ 
  &$ \text {BIAS}^2( \boldsymbol{\hat{\theta}})$ & 0.00435 & 0.00473 & 0.00220  \\ 
  &$ \text {VAR}( \boldsymbol{\hat{\theta}})$ & 1.74437 & 1.27188 & 0.26033 \\ 
  &$ \text {MAPE}( \boldsymbol{\hat{\theta}})$ & 5.46586 & 2.38020 & 0.52188  \\ 
  %&MAPE(bp) & 0.00275 & 0.02023 & 0.01135 \\
  &ACCE(bp) & 0.00221 & 0.01779 & 0.01122 \\ 
   \hline
\end{tabular}
\caption{\small Results for the max-EM algorithm with Binary Segmentation (BS) initialization in one and two breakpoint regression models. The first model is a linear homoscedastic regression model with two covariates, the second model is a logistic model with intercept and one covariate and the third model is a Weibull survival regression model with two covariates. The Mean Squared Error (MSE) of the estimated parameters, decomposed as the variance (VAR) plus squared bias ($\text{BIAS}^2$), 
                       along with the Mean Absolute Percentage Error (MAPE) of the estimated parameters and the ACCE of the estimated breakpoints are provided. The values of the true parameters can be found in Table~\ref{tab:param_setting_regression}.} 
\label{tab:reg_models}
\end{table}

\subsection{Implementation of the breakpoint tests in regression models}\label{ssec:bpDetectionSimu3}

In this section, we consider the statistical test developed in Section~\ref{sec:test} for the one breakpoint situation. This test is based on a permutation implementation where Theorems~\ref{th:likeli_test} and~\ref{th:likeli_test_small} are used for the computation of the distribution of the statistical test under $H_0$. The idea is simple: we randomly shuffle the order of the data $B=1,000$ times, and we consider that each shuffled sample is a realization of the test statistic. This realization is calculated using the approximations developed in Theorems~\ref{th:likeli_test} and~\ref{th:likeli_test_small} and therefore the max-EM algorithm does not need to be run. In practice, once this step has been performed, the p-value of the test can be computed by simply comparing the observed value of the statistical test on the original sample (using again Theorems~\ref{th:likeli_test} and~\ref{th:likeli_test_small}) with the distribution of the statistical test under $H_0$ obtained with the permutation implementation. By construction, the statistical test is automatically well calibrated under $H_0$: the rejection rate of the $\alpha$ level test under $H_0$ is equal to $\alpha$. However, it is of interest to investigate the power of the statistical test under various alternatives. This simulation experiment is conducted under the three regression models introduced in Section~\ref{ssec:bpDetectionSimu2}. In the linear model, Theorem~\ref{th:likeli_test} is used for samples larger than $100$, that is for $n_1=101,\ldots, 900$, in combination with Theorem~\ref{th:likeli_test_small} which is used for small samples (that is for $n_1< 100$ and $n_1\geq 900$). In the logistic and survival models, only Theorem~\ref{th:likeli_test} is used since the properties of the corresponding estimators are solely asymptotic. This amounts to constraining our test to detect a breakpoint for $n_1\geq100$ and $n_1< 900$ only. We start by considering the same parameter values as before (first scenario) and we then increase the difficulty in the segmentation detection in the second and third scenarios. The description of those scenarios with the corresponding values of the regression parameter values are given in Table~\ref{tab:param_setting_tests}. 

%For each scenario and each model, the statistical test under $H_0$ is computed using Theorems~\ref{th:likeli_test} and~\ref{th:likeli_test_small} by shuffling the data $M=1,000$ times for a sample of size $n=1,000$. In practice, once this step has been performed, the p-value of the test can be computed by simply comparing the observed value of the statistical test with this distribution. In our simulations, since we aim at evaluating the empirical power of the test, we 
%this step is always required in order to derive the distribution of the statistical test under $H_0$. Then, the observed value of the statistical test needs to be compared with this distribution by computing the p-value   The statistical test under $H_1$ is then computed on $M=1,000$ Monte-Carlo replications, using again Theorems~\ref{th:likeli_test} and~\ref{th:likeli_test_small}. 

\begin{figure}[H]
	\begin{center}
		\includegraphics[height=0.6\linewidth]{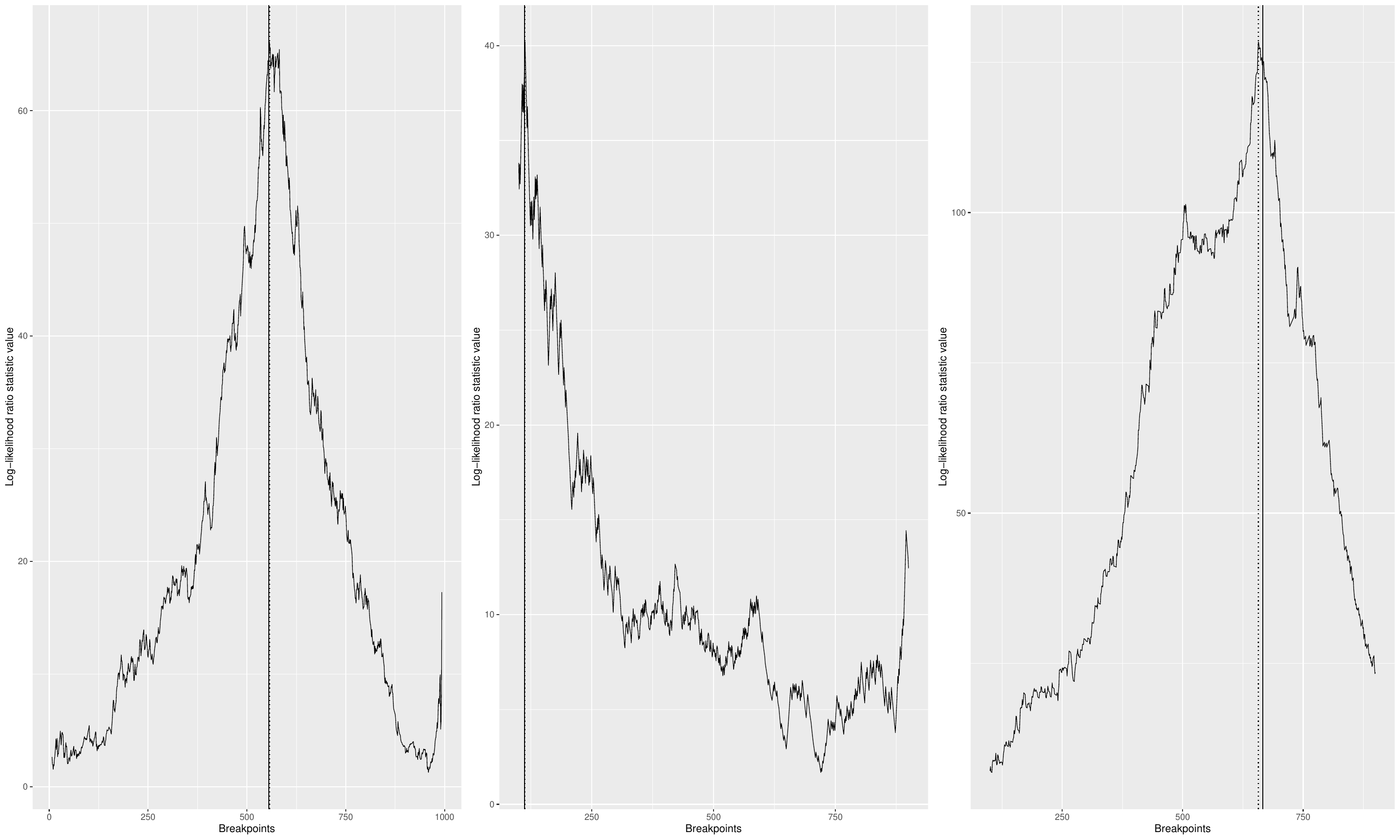}%height=0.5\linewidth
	\end{center}
\caption{\small{Log-likelihood ratio statistic for the test introduced in Section~\ref{sec:test}, in the linear (left panel), logistic (middle panel) and Weibull Cox (right panel) models. In each model, the true breakpoint is displayed as a plain vertical line and is equal to $555$, $112$ and $666$, respectively. The maximum of the log-likelihood ratio statistic is displayed as a dotted vertical line. The log-likelihood ratio statistic test is computed on a single sample using the approximations derived in Theorems~\ref{th:likeli_test} and~\ref{th:likeli_test_small}.}}   
	\label{fig:test_distribution_one}
\end{figure}

\begin{figure}[H]
	\begin{center}
		\includegraphics[height=0.9\linewidth]{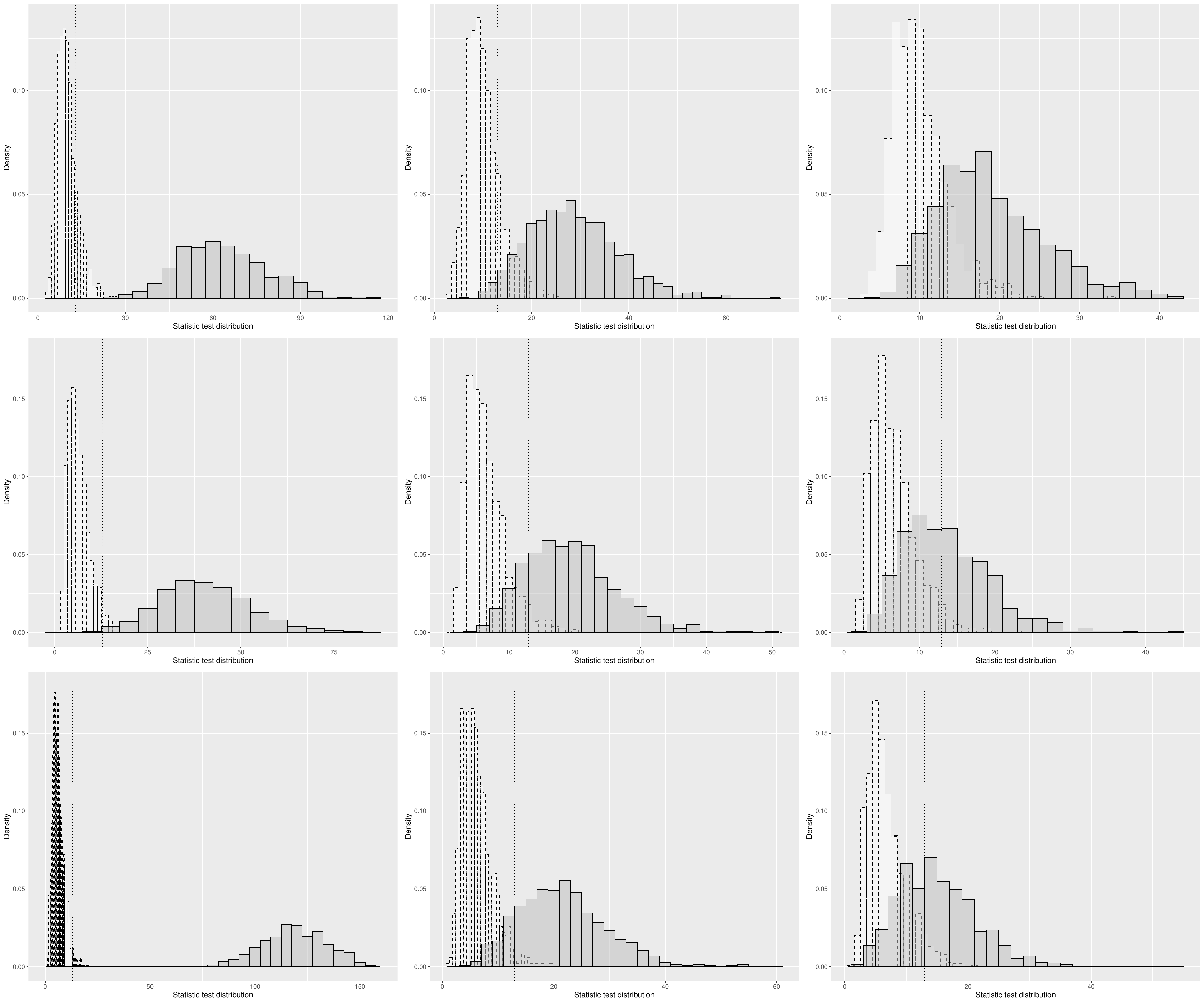}%height=0.5\linewidth
	\end{center}
\caption{\small{Distribution of the test statistics developed in Section~\ref{sec:test}, in the linear (top row), logistic (middle row) and survival (bottom row) models. The distribution under $(H_0)$ is represented by a white histogram with dotted contour line while the distribution under $(H_1)$ is represented by grey histogram with plain contour line. The $0.95$ quantile under $(H_0)$ is shown as a vertical dotted line. The three columns correspond to three scenarios with decreasing power (from left to right). Those scenarios are described in details in Section~\ref{ssec:bpDetectionSimu3} and Table~\ref{tab:param_setting_tests}. The log-likelihood ratio statistic test is computed using the approximations derived in Theorems~\ref{th:likeli_test} and~\ref{th:likeli_test_small}.}}   
	\label{fig:test_distribution}
\end{figure}

First, the log-likelihood ratio is computed on a single sample, for all possible breakpoint values and for all three models, in the first scenario. The value of the likelihood ratios with respect to the breakpoint values are displayed in Figure~\ref{fig:test_distribution_one}. On these samples, we clearly see that the maximum of the log-likelihood ratio is very close to the true value which is represented in dotted vertical lines in the figure. Then, the histograms of the statistical test are displayed in Figure~\ref{fig:test_distribution} in all situations, based on $M=1,000$ Monte-Carlo replications. The more the distribution under $H_1$ is far from the distribution under $H_0$, the more powerful the test is. For reference, the empirical $0.95$ quantile of the distribution under $H_0$ is shown as a vertical dotted line in order to visualize the power of the test for a $5\%$ level test. We clearly see that the power of the tests decreases as the distribution of the test statistic between $H_0$ and $H_1$ gets more similar (from left to right). For the linear model, the rejection rate under a $5\%$ level test is equal to $1$, $0.903$, $0.577$ for the left, middle and right panels, respectively. For the logistic model, the rejection rate under a $5\%$ level test is equal to $0.998$, $0.851$, $0.558$ for the left, middle and right panels, respectively. For the survival model, the rejection rate under a $5\%$ level test is equal to $1$, $0.928$, $0.601$ for the left, middle and right panels, respectively. Of importance, the permutation method is extremely fast to implement due to our approximations in Theorems~\ref{th:likeli_test} and~\ref{th:likeli_test_small}. For illustration, the computation of the $M=1,000$ samples used to derive the empirical distribution of the statistical test under $H_0$ is achieved in $18$ seconds on average, over all three scenarios, on a typical personal computer with 32Go of RAM.%statistical%, based on Theorem~\ref{th:likeli_test}

\begin{table}[ht]
\centering
%\resizebox{\textwidth}{!}{
\begin{tabular}{|lr|cc|cc|cc|}
  \hline
&&\multicolumn{2}{c|}{First scenario} & \multicolumn{2}{c|}{Second scenario}& \multicolumn{2}{c|}{Third scenario}\\
%\hline 
& & $\theta_1^*$& $\theta_2^*$& $\theta_1^*$& $\theta_2^*$ & $\theta_1^*$ & $\theta_2^*$   \\
 \hline
Linear &Intercept&1.00 & 2.00 &1.00 & 1.00 &1.00& 1.00 \\ 
($\sigma=2.5)$ & cov. effect 1&11.40 & 12.30 &11.00 & 12.30 &11.40&12.30 \\
 & cov. effect 2&0.60 & 0.10 &0.10 & 0.10 &0.10&0.10 \\
Logistic &Intercept&-1.10 & 0.50 &0.50 & 0.50 &0.50&0.50  \\ 
 & cov. effect &0.60 & -0.20 &1.20 & -0.20 &0.80 & -0.20  \\
 Survival &Intercept&2.00 & 2.50 &2.00 & 2.00 &2.00&2.00 \\ 
  & scale &1.70 & 1.98 &1.70 & 1.70 &1.70&1.70  \\ 
 & cov. effect 1&3.00 & 3.90&3.10 & 3.90 &3.30 & 3.90 \\ 
  & cov. effect 2&4.20 & 4.90&4.90 & 4.90 &4.90&4.90  \\ 
   \hline
\end{tabular}%}
\caption{\small Parameter values in the regression simulation scenarios for the statistical tests. A linear, logistic and survival models are studied in the one breakpoint setting. The linear model is homoscedastic with an error standard deviation equal to $2.5$ in all three scenarios. The breakpoints are equal to $553$, $112$ and $666$ in the linear, logistic and survival models, respectively.}%The data where simulated following a linear regression model with homoscedastic error, five breakpoints and one mean parameter in each segment 
\label{tab:param_setting_tests}
\end{table}

%
%\begin{figure}[H]
%	\begin{center}
%		\includegraphics[height=0.9\linewidth]{Test_Distributions.pdf}%height=0.5\linewidth
%	\end{center}
%\caption{Distribution of the test statistics developed in Section~\ref{sec:test}, in the linear (top row), logistic (middle row) and survival (bottom row) models. The distribution under $(H_0)$ is represented by a white histogram with dotted contour line while the distribution under $(H_1)$ is represented by grey histogram with plain contour line. The $0.95$ quantile under $(H_0)$ is shown as a vertical dotted line. The three columns correspond to three scenarios with decreasing power (from left to right). Those scenarios are described in details in Section~\ref{ssec:bpDetectionSimu3} and Table~\ref{tab:param_setting_tests}. The log-likelihood ratio statistic test is computed using the approximations derived in Theorems~\ref{th:likeli_test} and~\ref{th:likeli_test_small}.}   
%	\label{fig:test_distribution}
%\end{figure}

%

\section{Applications}\label{sec:realdata}

%=========    =========
\subsection{Tendency breakpoint detection on the bike sharing dataset}
\label{ssec:bpDetectionRealData}

\begin{figure}[htbp]
	\begin{center}
		\includegraphics[height=0.5\linewidth]{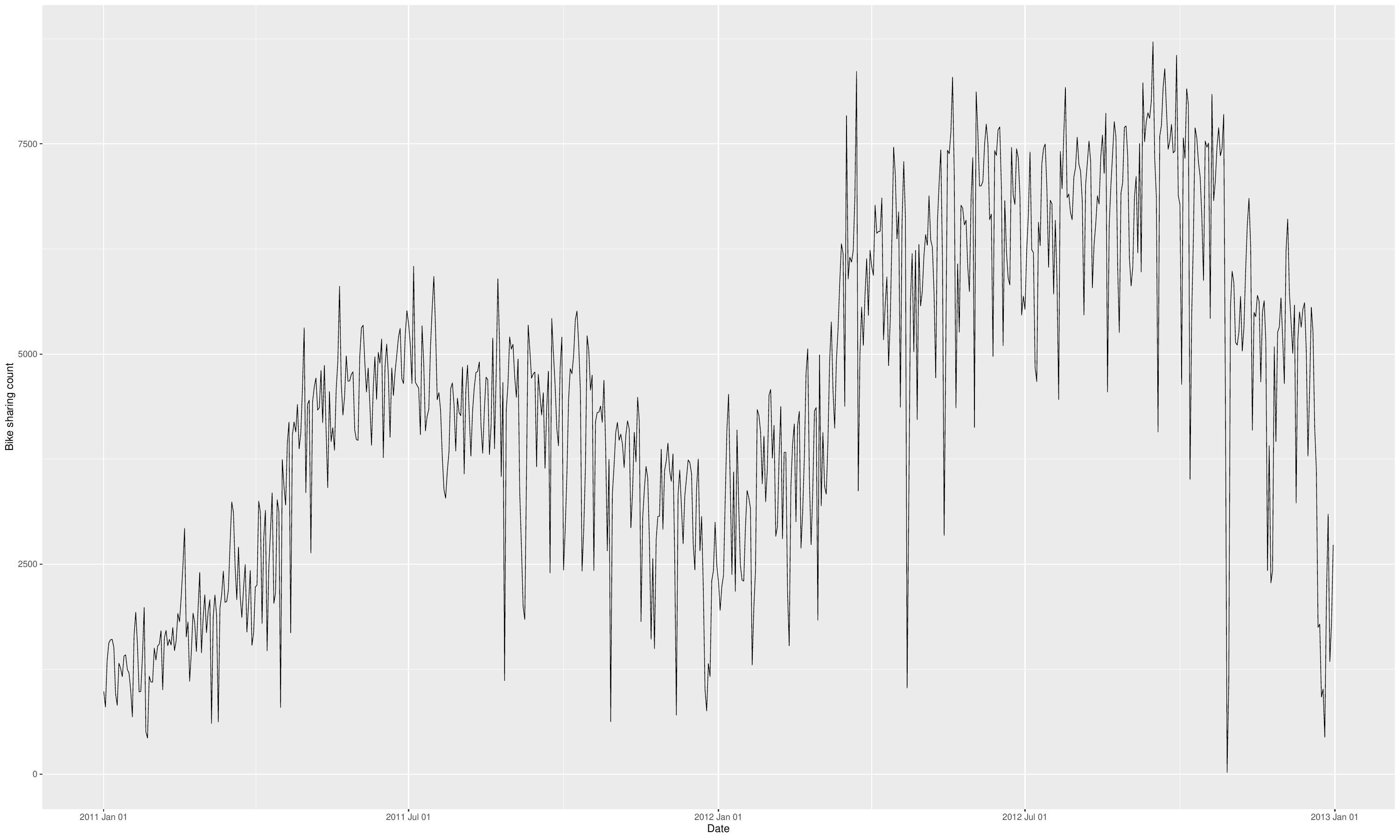}
	\end{center}
\caption{\small{Time series of bike sharing counts. The data are reported daily, from January 1, 2011 until December 31, 2012.}}   
	\label{fig:Bike_Sharing_ts}
\end{figure}

\begin{figure}[H]
	\begin{center}
		\includegraphics[height=0.5\linewidth,width=1\linewidth]{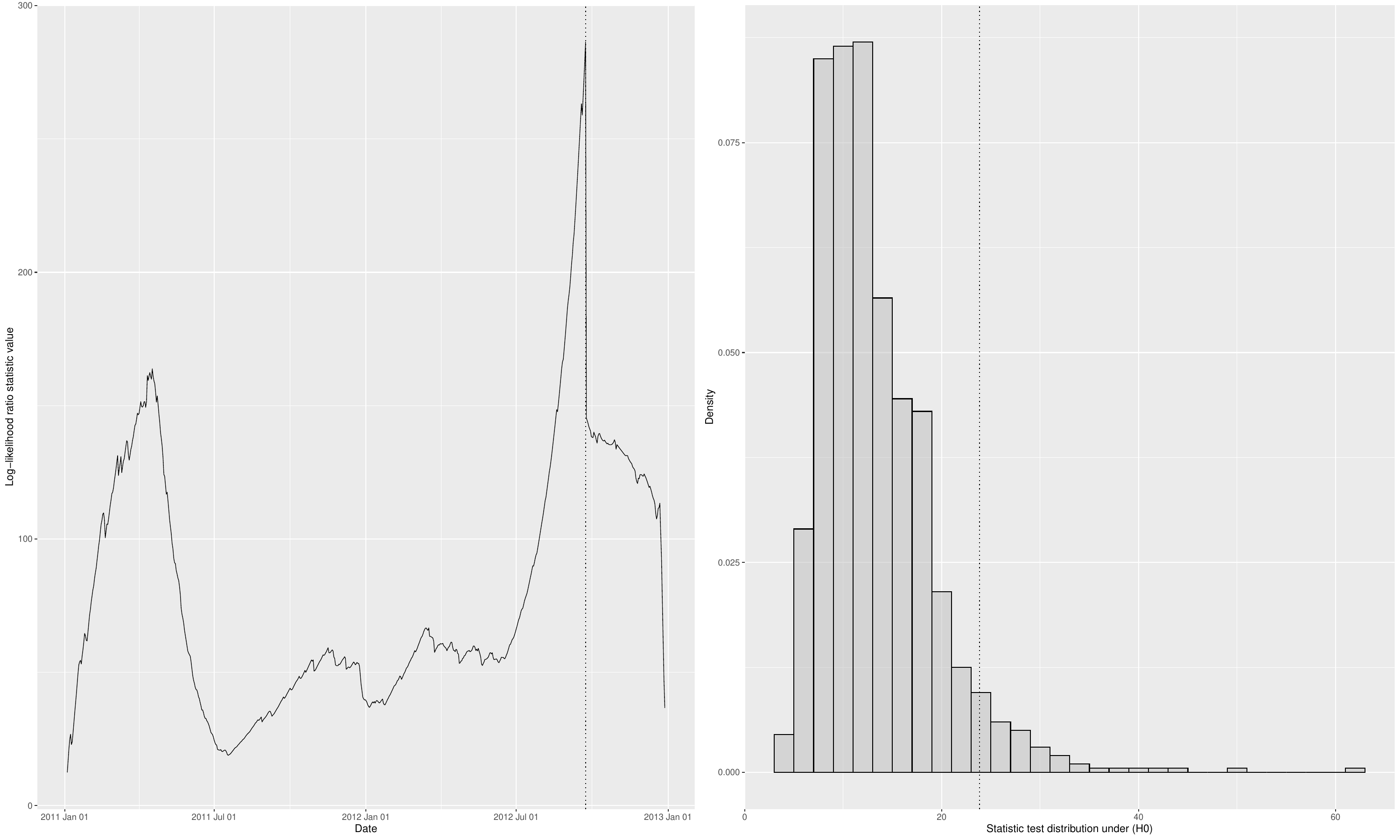}%height=0.5\linewidth
	\end{center}
\caption{\small{Statistical test for the one breakpoint detection problem for the bike sharing counts dataset. Left panel: log-likelihood ratio statistic computed on the original data. The maximum of the statistic is displayed as a vertical dotted line and equals $286.42$. Right panel: distribution of the test statistic under $H_0$. The distribution is obtained from $1,000$ permutations of the data. The $0.95$ quantile is shown as a dotted line. For both plots, the approximated expression of the log-likelihood ratio statistic was derived from Theorem~\ref{th:likeli_test}. The value of the test statistic obtained in the left panel ($286.42$) corresponds to a p-value equal to $0$.}}   %he approximated expression of the log-likelihood ratio statistic given in computed on
	\label{fig:LL_ratio_test}
\end{figure}

In this section we study the bike sharing dataset, available online on the UCI website. This dataset comprises the daily counts of the number of total rental bikes in a city from January 1, 2011 until December 31, 2012. It contains a total of $731$ values ($365$ in 2011 and $366$ in 2012). The time series is displayed in Figure~\ref{fig:Bike_Sharing_ts}. The aim is to study the trend of this time series and to detect change of trends with respect to the date. For that purpose, we use a simple linear regression model with intercept and the date as the only covariate. In the breakpoint analysis, we assume the model is homoscedastic, that is the variance of the residuals is the same in all segments. We start by performing the one breakpoint test. Using Theorem~\ref{th:likeli_test}, we compute the test statistic on those data and we simulate the test statistic under $H_0$ based on $B=1,000$ random permutations of the data. The results are shown in Figure~\ref{fig:LL_ratio_test}. The log-likelihood ratio statistic computed on the data is displayed on the left panel. We observe that the maximum is attained in $23$ September $2012$ and equals $286.42$. The empirical distribution under $H_0$ is displayed on the right panel with the $0.95$ empirical quantile represented as a vertical dotted line. We clearly see that, under $H_0$, the test statistic takes much lower values than $286.42$ and therefore the test is extremely significant with a p-value equal to $0$. When looking at the log-likelihood ratio statistic (on the left panel) we observe many other local maximums which have a value quite large as compared to the values taken by the test statistic under $H_0$. This suggests that the data may contain more breakpoints.

\begin{figure}[H]
	\begin{center}
		\includegraphics[height=0.5\linewidth,width=1\linewidth]{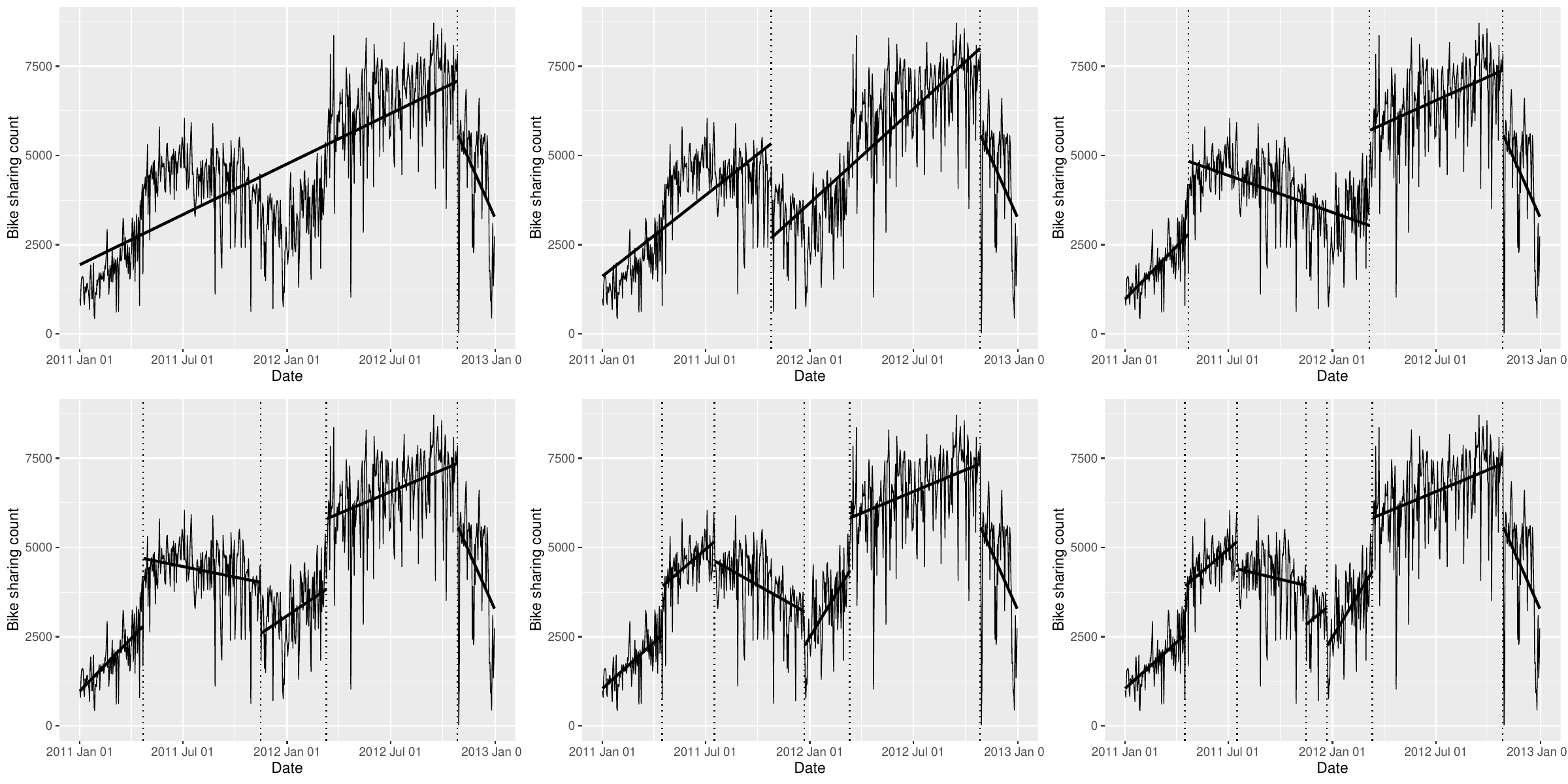}%height=0.5\linewidth
	\end{center}
\caption{\small{Tendency breakpoint detection and piecewise linear models implemented on the bike sharing dataset with various number of breakpoints, ranging from $1$ to $6$. The minimal value of the BIC is obtained for the $5$ breakpoint model.}}   %he approximated expression of the log-likelihood ratio statistic given in computed on
	\label{fig:Bp_Bike}
\end{figure}

We then apply our max-EM algorithm to the data, with a number of breakpoints ranging from $1$ to $6$. In Table~\ref{tab:bike_param}, we present the results of the different analyzes with the values of the estimated slopes and the value of the BIC computed using the expression introduced in Section~\ref{ssec:bicInferNBbp}. The values of the intercepts along with the dates at which the breakpoints occur can be found in Supplementary Materials. The plots of the linear models derived from these estimated parameters is also displayed in Figure~\ref{fig:Bp_Bike}. Up to five breakpoints, as the number of breakpoints increases, we clearly see an improvement in the data fitting, with very different values of slopes in two consecutive segments. On the contrary, in the six breakpoints model, the third and fourth breakpoints occur over a short period of time (2011-11-15 and 2011-12-22) with a change of slope sign ($-3.85$ and $12.73$) that does not seem to fit the data. Looking at the BIC value, it turns out that the five segments model is preferred over the six breakpoints model which is in agreement with Table~\ref{tab:bike_param} and Figure~\ref{fig:Bp_Bike}.% , in particular they are of the same sign, with few data in the segment five% the slopes between the segments four and five are similar ($12.7$ and $26.3$, respectively)

\begin{table}[ht]
\centering
\resizebox{\textwidth}{!}{
\begin{tabular}{|l|c|c|c|c|c|c|c|c|}
  \hline
bp & \multicolumn{7}{c|}{Slope values}& \multicolumn{1}{c|}{BIC}\\
% & 1 & 2 & 3 & 4 & 5 & 6 & 7 & 8 & 9 & 10 & 11 & 12 & 13 & 14 & 15 \\ 
  \hline
0 &  5.7688 &  &  &  &  &  &  & 12791.2900 \\ 
  1 &  7.7393 & -35.5764 &  &  &  &  &  & 12599.4121 \\ 
  2 &  12.5053 & 14.3050 & -35.5764 &  &  &  &  & 12411.8734 \\ 
  3 &  16.3069 & -5.6481 & 7.1842 & -35.5764 &  &  &  & 12193.2309 \\ 
  4 &  16.3069 & -3.2393 & 10.7407 & 6.7402 & -35.5764 &  &  & 12154.7065 \\ 
  5 &  14.2500 & 13.6033 & -8.9810 & 26.3382 & 6.6382 & -35.5764 &  & \bf{12149.2600} \\ 
  6 &  14.2500 & 13.6033 & -3.8540 & 12.7326 & 26.3382 & 6.6382 & -35.5764 & 12150.7580 \\ 
   \hline
\end{tabular}}
\caption{Estimated slope values obtained from the max-EM algorithm with the bike sharing counts dataset. The daily counts of shared bikes is modeled using a piecewise linear regression with respect to the dates in different models ranging from $0$ to $6$ breakpoints. The BIC is also reported in the last column.} %The data come from the bike sharing counts dataset and the daily counts of shared bikes is modelled using a piecewise linear regression with respect to the dates.} 
\label{tab:bike_param}
\end{table}

\subsection{Heterogeneity of the effect of fasting blood sugar on heart disease}
\label{ssec:bpHeteroHeart}

In this second real data application we study the heart disease dataset available on the UCI website. On this dataset of size $n=303$, the goal is to detect an heterogeneity in the effect of fasting blood sugar (fbs) on the risk of developing a heart disease ($54.46\%$ of the patients have a diagnostic of heart disease). In order to do so, we use the following continuous covariates: age, resting blood pressure on admission to the hospital (trestbps, in mm per Hg), cholesterol (chol, in mg per dl), maximum heart rate achieved (thalach) and ST depression induced by exercise relative to rest (oldpeak). Those covariates are used to construct a ``proximity space'' which allows us to order the individuals. Then we apply the max-EM algorithm for the logistic regression model where the outcome variable is the diagnostic of heart disease ($1$ yes, $0$ no) and the only covariate is fbs. This covariate is binary, with value $1$ when the fasting blood sugar exceeds $120$ mg/dl and $0$ when it is below this threshold.
The idea behind the construction of the proximity space is to find an order of individuals where two individuals whose ranks are close (respectively, far) to each other should be similar (respectively, different) in terms of covariates. To do so, we fit a principal curve~\citep[see][]{hastie1989principal} and we project the individuals on this curve. This is done using the \texttt{principal\_curve} function from the \texttt{princurve} R package. When the principal curve algorithm has converged, the order of individuals is obtained from the location on the curve (which is a space of dimension $1$) and we apply the max-EM algorithm to detect possible breakpoints.%theThe aim of the study is to detect heterogeneity on the effect of fbs on heart disease through % and reciprocally

\begin{figure}[H]
	\begin{center}
		\includegraphics[height=0.5\linewidth,width=1\linewidth]{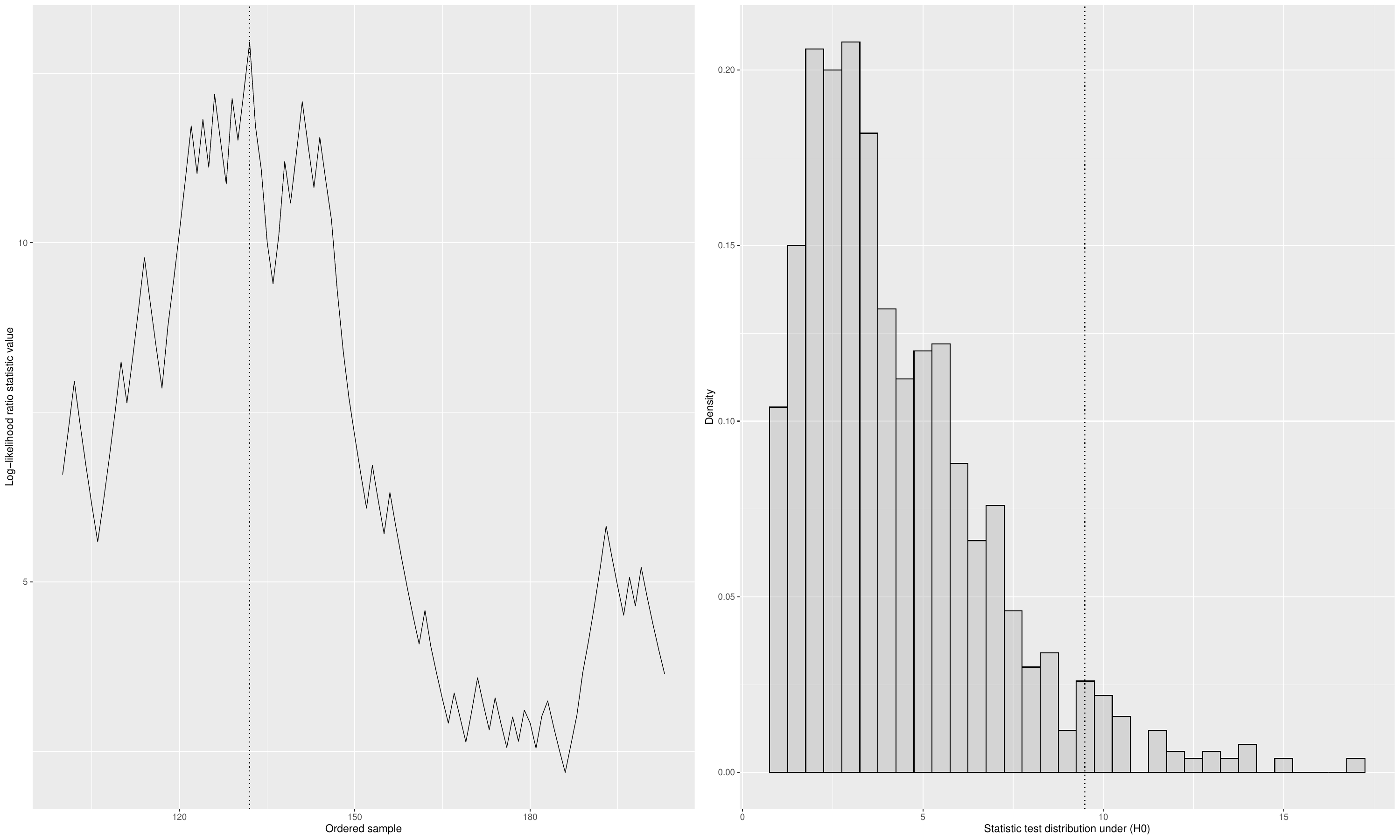}%height=0.5\linewidth
	\end{center}
\caption{\small{Statistical test for the one breakpoint detection problem for the heart disease dataset. Left panel: log-likelihood ratio statistic computed on the data ordered from the proximity space (this space was constructed from a principal curve fit on the covariates) based on a logistic regression model. The maximum of the statistic is displayed as a vertical dotted line and equals $12.96$. Right panel: distribution of the test statistic under $H_0$. The distribution is obtained from $1,000$ permutations of the data. The $0.95$ quantile is shown as a dotted line. For both plots, the approximated expression of the log-likelihood ratio statistic was derived from Theorem~\ref{th:likeli_test}. The value of the test statistic obtained in the left panel ($12.96$) corresponds to a p-value equal to $0.011$.}}   %he approximated expression of the log-likelihood ratio statistic given in computed on
	\label{fig:LL_ratio_test_heart}
\end{figure}

Before implementing the max-EM algorithm, we start by the test statistic for the one breakpoint scenario. In Figure~\ref{fig:LL_ratio_test_heart}, the log-likelihood ratio statistic is displayed on the left-panel for the ordered data with the maximum attained at the value $12.96$. On the right-panel, the distribution of the test statistic under $H_0$ is obtained based on Theorem~\ref{th:likeli_test} with the $0.95$ empirical quantile (equal to $9.48$) represented as a vertical dotted line. The p-value is simply the probability that this density is greater than $12.96$ and it equals $0.011$. The test is therefore highly significant and suggests that the effect of fbs is heterogeneous according to a breakpoint on the principal curve space. Since the ordering of individuals on this space was obtained based on covariates proximity, this suggests an interaction effect of covariates/fbs on the diagnosis of heart disease. Next, the max-EM algorithm is implemented with different breakpoint models. The result of the BIC along with the odds ratios for fbs on the diagnosis of heart disease are displayed in Table~\ref{tab:heart_disease}. We observe that the model with minimum value for the BIC is the one breakpoint model for which the odds ratios in the two segments are equal to $0.56$ and $1.12$, respectively. This means that fbs has a strong protective effect for individuals in segment $1$ and a slightly worsening effect for individuals in segment $2$. In the one breakpoint model, the two segments are of size $n=132$ and $n=171$, respectively.

\begin{table}[ht]
\centering
%\resizebox{\textwidth}{!}{
\begin{tabular}{|l|c|c|c|c|}%c|c|c|c|
  \hline
bp & \multicolumn{3}{c|}{Odds ratios for fbs}& \multicolumn{1}{c|}{BIC}\\
% & 1 & 2 & 3 & 4 & 5 & 6 & 7 & 8 & 9 & 10 & 11 & 12 & 13 & 14 & 15 \\ 
  \hline
0 &  0.8540 &    &  &  428.8278 \\ %&  &  &  &
  1 &  0.5611 & 1.1209 &    & \textbf{427.1403} \\ %&  &  &  &
  2 &  0.5611 & 0.9698 & 4.5000   & 432.5396 \\ %&  &  &  &
 % 3 &   & - &  & - &  &  &  & 437.8522 \\ 
  %4 &   & - &  &  & - &  &  & 441.2868 \\ 
   \hline
\end{tabular}%}
\caption{Estimated odds ratios obtained from the max-EM algorithm with the heart disease dataset. The odds for fasting blood sugar on diagnosis of heart disease is modelled using logistic regression based on a proximity space constructed from the principal curve of $5$ different covariates. Models ranging from $0$ to $2$ breakpoints are presented. The BIC is also reported in the last column.} %The data come from the bike sharing counts dataset and the daily counts of shared bikes is modelled using a piecewise linear regression with respect to the dates.} 
\label{tab:heart_disease}
\end{table}

\begin{figure}[H]
	\begin{center}
		\includegraphics[height=0.5\linewidth,width=1\linewidth]{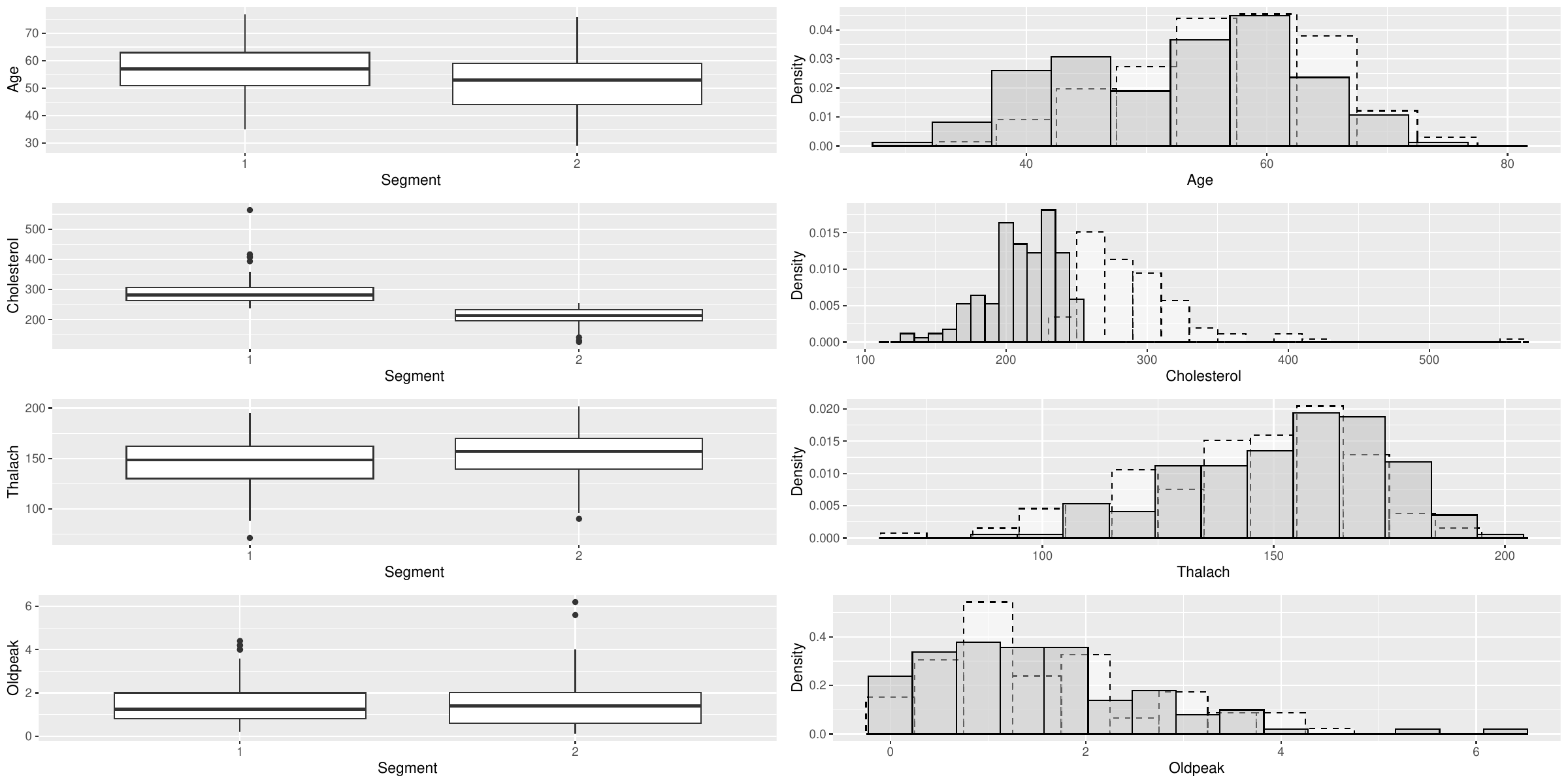}%height=0.5\linewidth
	\end{center}
\caption{\small{Univariate distributions of the covariates used to construct the proximity space between the two segments. The breakpoint, and therefore the two segments, were obtained from the max-EM algorithm. Left panel: boxplots of the covariates. Right panel: histograms of the covariates in white with dotted contour lines and in grey with plain contour line for the data in segment 1 and segment 2, respectively.}}   %he approximated expression of the log-likelihood ratio statistic given in computed on
	\label{fig:Heart_Cov}
\end{figure}

In order to investigate what can cause the odds ratios to be twice as big in segment 2 as compared to segment 1, we have also compared the distributions of the covariates in the two segments. We present, in Figure~\ref{fig:Heart_Cov}, the univariate distributions of the $5$ covariates. Since the oldpeak variable has a lot of zeros (which means the patient had no ST depression), the distribution of this variable is for the positive values only (for reference, there are a total of $99$ individuals with a value of oldpeak equal to $0$ in both segments, which correspond to $30.3\%$ and $34.5\%$ of oldpeak values equal to $0$ in segments 1 and 2, respectively). We observe that the main covariate that distinguishes the two segments is cholesterol with much lower values in segment 2 as compared to segment 1 (median with interquartile range equals $282 \,[264,307]$ and $214 \, [197,232]$ in segments 1 and 2, respectively). Then, individuals in segment 2 tend to be younger ($57 \, [51,63]$ in segment 1 and $53 \, [44,59]$ in segment 2), with a higher value of thalach ($148.5 \, [130,162]$ in segment 1 and $157 \, [140.5,170]$ in segment 2). Regarding the oldpeak variable, there are more patients with no ST depression in segment 2, but among those who had ST depression, the ST depression value tends to bee slightly larger in segment 2 than in segment 1 ($1.25\, [0.8,2]$ in segment 1 and $1.4 \, [0.6,2.02]$ in segment 2). The correlation between all pair of variables was also studied and compared between each segment. We present all the pairwise correlation values in Table~4 in Supplementary Material along with the scatter plots of some of the variables in Figure~1. Focusing on only the strongest associations between pair of variables, we see that: age and thalach are negatively correlated with a correlation equal to $-0.253$ and $-0.461$ in segments 1 and 2, respectively; age is positively correlated with trestbps (it is equal to $0.209$ and $0.301$ in segments 1 and 2, respectively); thalach is positively correlated with cholesterol (it is equal to $0.187$ and $0.228$ in segments 1 and 2, respectively) and oldpeak is positively correlated with trestbps (it is equal to $0.253$ and $0.151$ in segments 1 and 2, respectively).

\section{Discussion and perspectives}

%================================ Annexe ====================================
%\addresseshere

In this work we presented a new method for breakpoint detection in regression modeling. Our method, called max-EM, which combines the CEM algorithm with HMM, is an extension of previous approaches on the topic based on the standard EM algorithm. We showed that it is tailored to the breakpoint detection problem: when the targeted likelihood is a function of both the regression parameters and breakpoint locations, we proved that each iteration of the max-EM algorithm increases this likelihood. We also presented two strategies for the initialization of the algorithm and we proposed to use the standard BIC in practice to find the correct number of breakpoints. Finally, a new statistical test for the one breakpoint situation based on the likelihood ratio for all possible segments has been studied: we established an asymptotic approximation that allows to compute this test in an efficient and fast way.

%Our algorithm
As compared to the GFPOP algorithm, our method does not provide an exhaustive exploration of all possible segmentations but rather, is based on statistical models and aims at increasing the likelihood at each iteration. Using our initialization methods, our method becomes more stable and seems to be able to reach global maximums. It is extremely fast, even though the initialization step requires to run the algorithm several times. Our simulation experiments tend to favor the BS initialization over the FL initialization, in terms of computation time - accuracy balance. Importantly, our method can consider a very wide variety of regression models, a feature that is not possible using dynamic programming. In our simulation settings and in the analyzes of real data, we indeed considered linear, logistic, Poisson, and AFT regression models. We showed that in all these regression settings, with a number of breakpoints ranging from $1$ to $5$, our method was extremely performant, both in terms of breakpoint detection and parameters estimation. The statistical test was also studied under various regression models. It showed a correct rejection rate under the null hypothesis and a strong power under some alternative hypothesis. This was illustrated on the two studied datasets where the test was highly significant, in particular for the bike sharing dataset. Using the BIC to choose the correct number of breakpoints seemed also to be a powerful approach. In the two datasets we found relevant segments where the distribution of the data was clearly different between segments. Those applications showed the versatility of our approach. In the bike sharing dataset, it was used to detect change of trends in the number of total rental bikes with respect to the date. In the heart disease data, it was combined with the construction of principal curves to construct a proximity space on the covariates. This proximity space was then used to define the order of individuals and combined with the max-EM algorithm this enabled us to detect different effects of fasting blood sugar on the occurrence of heart disease. The segments were composed of covariates with similar values among segments and different values between segments. Analyzing the distribution of the covariates inside the two segments, the whole procedure enabled us to detect complex interactions between the effect of fasting blood sugar and the other covariates on the occurrence of heart disease. 
%Combining our method with 

%lead to

%Our max-EM algorithm also allows to consider settings where some regression parameters are imposed to be in common in all the segments. This was studied in the simulation section for the homoscedastic regression model. This is also an attractive feature in the context of censored data, where one might want to use the popular Cox model. When the variable of interest is a time variable, it might be relevant to detect changes in terms of hazard ratios of a covariate of interest between segments and to keep the baseline common to all segments.  %modeling approach

A setting that we did not investigate in this work occurs when some regression parameters are imposed to be shared between segments. This is an attractive modeling approach, however the current method does not support this feature. This is due to the fact that the parameter update step of the algorithm simply consists in estimating the parameters in each segment (thus leading to different estimations per segments). In our simulations, we considered scenarios where some of the parameters are common over the segments: we simulated data following the homoscedastic linear regression and the homoscedatic AFT model. In those settings, our method did not take advantage of the homoscedastic structure of the data but still provided accurate parameter estimations. However, it would be of interest to develop a method that explicitly incorporates this feature in the estimation method. In particular, this would be extremely relevant in the context of censored data, where one wants to use the popular Cox model. When the variable of interest is a time variable, it might be relevant to detect changes in terms of hazard ratios of a covariate of interest between segments and to keep the baseline common to all segments. This modeling option would need further work, both for the max-EM algorithm and for the one breakpoint statistical test. Regarding the test, this would be particularly relevant as our current approach might detect heterogeneity due to baseline differences among segments, when one might only be interested in changes in the covariate effect. This is left to future research work.

%(corresponding to all possible breakpoints)  
 
%established

\section*{Acknowledgement}

The authors warmly thank Guillem Rigaill and Vincent Runge for our fruitful discussions on the GFPOP algorithm. This work is part of the project entitled ``A new method for the detection of gene-environment interactions in cancer studies'' and was funded by the Ligue Nationale Contre le Cancer (LNCC).
%We thank the reviewers for their constructive criticisms and comments that have helped improve the paper.

%\clearpage
%\newpage

\renewcommand\thesection{\Alph{section}}
\setcounter{section}{0}
\renewcommand{\theequation}{\Alph{section}.\arabic{equation}}

\section{Appendix}

%In order to prove Theorem~\ref{th:likeli_test} and~\ref{th:likeli_test_small} we first recall a classical result for likelihood based models.
%
%\begin{lemma}\label{lem:likeli}
%Let $n, n_1\in\mathbb N^*$, such that $n>n_1$ and $n_1\to\infty$, $n-n_1\to\infty$. Then, under standard assumptions for maximum likelihood theory, 
%\begin{align*}
%\tilde\ell_n(\hat\theta_0) & = \tilde\ell_n(\theta_0^*) +\nabla \tilde\ell_n(\theta_0^*)^{\top}\left(-\nabla^2\tilde\ell_n(\theta_0^*)\right)^{-1}\nabla \tilde\ell_n(\theta_0^*)+o_{\mathbb P}(1)
%\end{align*}
%and
%\begin{align*}
%\ell_n(\hat\theta_1,\hat\theta_2) & = \ell_n(\theta_0^*,\theta_0^*) +\frac 12\nabla \ell_n(\theta_0^*,\theta_0^*)^{\top}\left(-\nabla^2\ell_n(\theta_0^*,\theta_0^*)\right)^{-1}\nabla \ell_n(\theta_0^*,\theta_0^*)+o_{\mathbb P}(1).
%\end{align*}
%\end{lemma}

\subsection{Proof of Proposition~\ref{prop:cvgce_EM}}

At the $(m+1)$th step, we have for $k={R_{i}^{\text max}}^{(m+1)}$, for all $k'=1,\ldots,K$, for all $i=1,\ldots,n$, $F_{i}^{\text{max}}(k ; {\boldsymbol{\theta}}^{(m)}) B_{i}^{\text{max}}(k;{\boldsymbol{\theta}}^{(m)})\geq F_{i}^{\text{max}}(k' ; {\boldsymbol{\theta}}^{(m)}) B_{i}^{\text{max}}(k';{\boldsymbol{\theta}}^{(m)})$. From Equation~\eqref{eq:maxFB} and the definition of $\ell_n$ in Equation~\eqref{eq:likelihood}, we therefore have
\begin{align*}
\ell_n\left(\boldsymbol{\theta}^{(m)};n^{(m+1)}_{1:(K-1)}\right)\geq \ell_n\left(\boldsymbol{\theta}^{(m)};n^{(m)}_{1:(K-1)}\right).
\end{align*}
Now, from the M-step, $\boldsymbol\theta^{(m+1)}$ is the maximizer of $\ell_n\left(\boldsymbol{\theta};n^{(m+1)}_{1:(K-1)}\right)$ and 
%\begin{align*}
%\theta^{(m+1)}=\underset{\boldsymbol{\theta}}{\operatorname{argmax}} \;\ell_n\left(\boldsymbol{\theta};n^{(m+1)}_{1:(K-1)}\right),
%\end{align*}
consequently
\begin{align*}
\ell_n\left(\boldsymbol{\theta}^{(m+1)};n^{(m+1)}_{1:(K-1)}\right)\geq \ell_n\left(\boldsymbol{\theta}^{(m)};n^{(m+1)}_{1:(K-1)}\right).
\end{align*}
This proves that the sequence $\left(\ell_n\left(\boldsymbol{\theta}^{(m)};n^{(m)}_{1:(K-1)}\right)\right)_{m\geq 1}$ is increasing. Since there is a finite number of partition of the segments $R_{1:n}$ under the contraint $R_n=K$ and since $e_i(k;\theta_k)$ is bounded, the log-likelihood $\ell_n\left(\boldsymbol{\theta}^{(m)};n^{(m)}_{1:(K-1)}\right)$ converges towards a finite value. Moreover the maximum is unique by assumption and as a consequence $\left(\boldsymbol{\theta}^{(m)};n^{(m)}_{1:(K-1)}\right)_{m\geq 1}$ converges towards a stationary point.%$\left(\boldsymbol{\theta}^{*};n^{*}_{1:(K-1)}\right)$.%, equal to $\ell_n\left(\boldsymbol{\theta}^{*};n^{*}_{1:(K-1)}\right)$.%concludes the proof.

\subsection{EM and max-EM algorithms}

%\subsubsection{E-step of the EM algorithm}
%
%\label{sssec:annexeEM}
%
%\begin{align*}
	%\mathbb{Q}(\theta|\theta^{\text{old}}) = \mathbb{E}_{[R| {\color{black}X} ; \theta^{\text{old}}]} [\log \mathbb{P}({\color{black}Y}, R| Z; \theta)] &= \sum_{R} \mathbb{P}(R| Y, Z; \theta^{\text{old}}) \log \mathbb{P}({\color{black}Y}, R| Z; \theta) \\
	% &{\color{red}= \dots}  \\ % Dans l'annexe : d�j� pr�sent� dans le mat�riel de l'article d'Olivier et Gr�gory; 
	% &{\color{red}= \dots}  \\
	% &{\color{red}= \dots}  \\
	% &{\color{red}=  \sum_{i = 1}^n \sum_{k = 1}^K  \log\{e_i(k ; \theta)\} \omega_i(k; \theta^{\text{old}})} % Attention : lors des sommes est invers� dans Celeux --> r�ponse Celeux regarde vraiment la vraisemblance dans l'intro alors qu'on est ici sur Q qui est diff�rent de la vraie vraisemblance.  
	%\end{align*}

\subsubsection{MAP in the E-step of the max-EM algorithm}

\noindent The $F_{i}^{\text{max}}(k ; \theta)$ and $B_{i}^{\text{max}}(k;\theta)$ can be combine to compute 
\begin{flalign*}
	F_{i}^{\text{max}}(k ; \theta) B_{i}^{\text{max}}(k;\theta) &\equiv  \underbrace{F_{i}^{\text{max}}(R_{i} = k;\theta) B_{i}^{\text{max}}(R_{i} = k;\theta)}_{ = \text{MAP in } R_{i}}.& \\
	%%&= {\color{orange}\max_{R_{-j}} \mathbb{P}(R_{j},R_{-j},D)} &\\
	%%&= \max_{R_{-j}} \prod_{i = 1}^{n} \phi_{i}(R_{i-1},R_{i})&  \\
	%&={\color{cyan}\underset{R_{1:i-1}}{\operatorname{max}} \mathbb{P}(R_1, \dots, R_{i-1}, R_i = k,X_{1:n}|\theta)} &\\
	&={\color{black}\underset{R_{1:i-1}, R_{(i+1):(n-1)}}{\operatorname{max}} \mathbb{P}(R_1, \dots, R_{i-1}, R_i = k,R_{i+1},\dots, R_{n} = K,X_{1:n} | \theta)}&
\end{flalign*}

\noindent \textit{Proof:} From
\begin{align*}
	F_{i}^{\text{max}}(k ; \theta) &= \max_{R_1, \dots, R_{i-1}} \mathbb{P}(R_{1:i-1}, R_i = k,{\color{black}X_{1:i}}| \theta),\\
	\text{ and } \\
	B_{i}^{\text{max}}(k;\theta) &= \max_{R_{i+1}, \dots, R_{n-1}} \mathbb{P}(R_{(i+1):(n-1)}, R_{n} = K, X_{(i+1):n}| R_i = k;\theta).
\end{align*}

%\begin{align*}
	%	F_{i}^{\text{max}}(k ; \theta) = \max_{R_1, \dots, R_{i-1}} \mathbb{P}(R_1, \dots, R_{i-1}, R_i = k,D_{1}, \dots, D_{i}),
	%\end{align*}
%and
%\begin{align*}
	%	B_{i}^{\text{max}}(k;\theta) = \max_{R_{i+1}, \dots, R_{n-1}} \mathbb{P}(R_{i+1}, \dots, R_{n},D_{i+1}, \dots, D_{n}| R_i = k). 
	%\end{align*}

We compute the product $F_{i}^{\text{max}}(k ; \theta)\times B_{i}^{\text{max}}(k;\theta)$ as  
\begin{align*}
	F_{i}^{\text{max}}(k ; \theta)\times B_{i}^{\text{max}}(k;\theta) & =\max_{R_1, \dots, R_{i-1}} \mathbb{P}(R_{1:i-1}, R_i = k,{\color{black}X_{1:i}})\\
	& \quad \times \max_{R_{i+1}, \dots, R_{n-1}} \mathbb{P}(R_{(i+1):(n-1)}, R_{n} = K,X_{(i+1):n}| R_i = k;\theta)
\end{align*}

\noindent Then, considering that
\begin{align*}
	&\mathbb{P}(R_{(i+1):(n-1)}, R_{n} = K, {\color{black}X_{(i+1):n}| R_i = k;\theta}) \\
	&=  \mathbb{P}(R_{(i+1):(n-1)}, R_{n} = K, {\color{black}X_{(i+1):n}| R_i = k,R_{1:(i-1)},X_{1:i};\theta})
\end{align*}
we obtain
\begin{align*}
	& \mathbb{P}(R_{1:i-1}, R_i = k,{\color{black}X_{1:i}}; \theta) \\
	& \times \mathbb{P}(R_{(i+1):(n-1)}, R_{n} = K, {\color{black}X_{(i+1):n}| R_i = k,R_{1:(i-1)},X_{1:i};\theta})\\
	& =\mathbb{P}(R_{(i+1):(n-1)}, R_{n} = K,X_{(i+1):n}, R_i = k,X_{1:i},R_{1:(i-1)}|\theta)
\end{align*}
Thus
\begin{align*}
	F_{i}^{\text{max}}(k ; \theta)B_{i}^{\text{max}}(k;\theta) & =\max_{R_{1:(i-1)}, R_{(i+1):(n-1)}}\mathbb{P}(R_{(i+1):(n-1)}, R_{n} = K,X_{1:n}, R_i = k,R_{1:(i-1)} | \theta)
\end{align*}

\subsection{Forward Backward and Max-Forward Max-Backward algorithms}

\subsubsection{Forward Backward algorithm in logarithmic scale}

\label{sssec:annexeFwBwLogScale}

In order to avoid the underflow problem, %(when very small quantities are computed directly), 
we factor the results into a logarithmic scale: %by rewritting:
\begin{align*}
	e_i(k; \theta) = \text{e}^{\ell_i} \tilde{e}_i(k; \theta), \hspace{0.7cm} F_{i}(k; \theta) = \text{e}^{L_i} \tilde{F}_{i}(k; \theta)   \hspace{0.3cm} \text{ and } \hspace{0.3cm}  B_{i}(k; \theta) = \text{e}^{M_i} \tilde{B}_{i}(k; \theta), 
\end{align*}
with
\begin{align*}
	\text{e}^{\ell_i} = \max_{k} e_i(k; \theta), \hspace{0.7cm} \text{e}^{L_i} = \max_{k} F_{i}(k; \theta)   \hspace{0.3cm} \text{ and } \hspace{0.3cm} \text{e}^{M_i} = \max_{k} B_{i}(k; \theta) , 
\end{align*}

\vspace{0.5cm}
From there, we find for the forward quantities (the same holds for the backward quantities):
\begin{align*}
	F_{i}(k; \theta) &= \sum_{j} F_{i-1}(j ; \theta) \pi(j,k) e_i(k; \theta) \hspace{0.2cm} \Leftrightarrow \hspace{0.2cm} \text{e}^{L_i}  \tilde{F}_{i}(k; \theta) = \text{e}^{L_{i-1} + \ell_i}  \sum_{j} \tilde{F}_{i-1}(j;\theta) \pi(j,k) \tilde{e}_i(k; \theta) 
\end{align*}

\vspace{0.5cm}

\noindent Therefore, 
\vspace{-0.1cm}
\begin{align*}
	L_i = L_{i-1} + \ell_i +\max_{k} \log \Big( \sum_{j} \tilde{F}_{i-1}(j;\theta) \pi(j,k) \tilde{e}_i(k; \theta) \Big)  \hspace{0.2cm} \text{ and }  \tilde{F}_{i}(k; \theta) \propto  \sum_{j} \tilde{F}_{i-1}(j;\theta) \pi(j,k) \tilde{e}_i(k; \theta) \hspace{0.2cm}
\end{align*}

\vspace{-0.2cm}
and, in the same way:
\vspace{-0.1cm}
\begin{align*}
	M_{i-1} = M_{i} + \ell_i +\max_{{\color{black}j}} \log \Big( \sum_{k}  \pi(j,k) \tilde{e}_i(k; \theta) \tilde{B}_{i}(k; \theta) \Big)  \hspace{0.2cm} \text{ and }  \tilde{B}_{i-1}(j;\theta) \propto  \sum_{k}  \pi(j,k) \tilde{e}_i(k; \theta) \tilde{B}_{i}(k; \theta) \hspace{0.2cm}
\end{align*}

\subsubsection{Max-Forward Max-Backward algorithm in logarithmic scale}

\label{sssec:annexeMaxFwMaxBwLogScale}

similarly with the forward backward algorithm, we find for the forward quantities (the same holds for the backward quantities):
\begin{align*}
	F_{i}^{\text{max}}(k ; \theta) &= \underset{j}{\operatorname{max}} \ F_{i-1}^{\text{max}}(j; \theta) \pi(j,k) e_i(k; \theta) \hspace{0.2cm} \Leftrightarrow \hspace{0.2cm} \text{e}^{L_i}  \tilde{F}_{i}^{\text{max}}(k;\theta) = \text{e}^{L_{i-1} + \ell_i}  \underset{j}{\operatorname{max}} \ \tilde{F}_{i-1}^{\text{max}}(j) \pi(j,k) \tilde{e}_i(k; \theta) 
\end{align*}

\vspace{0.1cm}

\noindent Therefore, 
\vspace{-0.1cm}
\begin{align*}
	L_i = L_{i-1} + \ell_i +\max_{k} \log \Big( \underset{j}{\operatorname{max}} \ \tilde{F}_{i-1}^{\text{max}}(j) \pi(j,k) \tilde{e}_i(k; \theta) \Big)  \hspace{0.1cm} \text{ and }  \tilde{F}_{i}^{\text{max}}(k;\theta) \propto  \underset{j}{\operatorname{max}} \ \tilde{F}_{i-1}^{\text{max}}(j) \pi(j,k) \tilde{e}_i(k; \theta) \hspace{0.2cm}
\end{align*}

\vspace{-0.1cm}

and, in the same way:
\vspace{-0.1cm}
\begin{align*}
	M_{i-1} = M_{i} + \ell_i +\max_{{\color{black}j}} \log \Big( \underset{k}{\operatorname{max}} \  \pi(j,k) \tilde{e}_i(k; \theta) \tilde{B}_{i}^{\text{max}}(k;\theta) \Big)  \hspace{0.0cm} \text{ and }  \tilde{B}_{i-1}^{\text{max}}(j) \propto  \underset{k}{\operatorname{max}} \  \pi(j,k) \tilde{e}_i(k; \theta) \tilde{B}_{i}^{\text{max}}(k;\theta) \hspace{0.2cm}
\end{align*}

\subsection{Statistical Tests: Theorem Proofs}

\subsubsection{Proof of Theorem~\ref{th:likeli_test}}

%The idea for the proof is to study the likelihood ratio test statistic for a fixed value of $n_1$ and then to take the maximum over those ratios. We first define, for a fixed value $n_1=\lfloor n\lambda_j \rfloor$, $j\in\{1,\ldots,J\}$,
We first recall that
\begin{align}\label{eq:0}
\ell_n(\theta_1,\theta_2)&=\sum_{i=1}^{n_1} \log\left(\mathbb P(X_i;\theta_1)\right)+ \sum_{i=n_1+1}^{n} \log\left(\mathbb P(X_i;\theta_2)\right),\nonumber\\
\tilde\ell_n(\theta)&=\sum_{i=1}^{n} \log\left(\mathbb P(X_i;\theta)\right)=\sum_{i=1}^{n_1} \log\left(\mathbb P(X_i;\theta)\right)+ \sum_{i=n_1+1}^{n} \log\left(\mathbb P(X_i;\theta)\right),
\end{align}
and $(\hat\theta_1,\hat\theta_2)=\argmax_{\theta_1,\theta_2}\ell_n(\theta_1,\theta_2)$, $\hat\theta_0=\argmax_{\theta}\tilde\ell_n(\theta)$. 
%Also, let $(\hat\theta_1,\hat\theta_2)=\argmax_{\theta_1,\theta_2}\ell_n(\theta_1,\theta_2)$ and $\hat\theta_0=\argmax_{\theta}\tilde\ell_n(\theta)$. 
It is clear that $\tilde\ell_n(\theta)=\ell_n(\theta,\theta)$ but it should be noted that the gradient and Hessian matrix for $\tilde\ell_n$ and $\ell_n$ are different even if they are evaluated at the same parameter value. In particular, $\nabla \ell_n(\theta_1,\theta_2)$ is a $2d$ dimensional vector where the first $d$ components contain the vector $\sum_{i=1}^{n_1}\nabla \log\left(\mathbb P(X_i;\theta_1)\right)$ and the last $d$ components contain the vector $\sum_{i=n_1+1}^{n}\nabla \log\left(\mathbb P(X_i;\theta_2)\right)$. The Hessian matrix $\nabla^2 \ell_n(\theta_1,\theta_2)$ is a $2d\times 2d$ matrix which can be decomposed as four $d\times d$ block matrices in the following way:
\begin{align*}
\nabla^2 \ell_n(\theta_1,\theta_2)=\begin{pmatrix}\sum_{i=1}^{n_1}\nabla^2 \log\left(\mathbb P(X_i;\theta_1)\right) & 0_{d\times d}\\ 0_{d\times d} & \sum_{i=n_1+1}^{n}\nabla^2 \log\left(\mathbb P(X_i;\theta_2)\right) \end{pmatrix}
\end{align*}

%. The first block contains the matrix $\sum_{i=1}^{n_1}\nabla^2 \log\left(\mathbb P(X_i;\theta_1)\right)$ and the lower right block contains the matrix $\sum_{i=1}^{n_1}\nabla^2 \log\left(\mathbb P(X_i;\theta_2)\right)$. The two outer diagonal blocks are composed of $0$. On the other hand, $\nabla \tilde\ell_n(\theta)$ is a $d$ dimensional vector and $\nabla^2 \tilde\ell_n(\theta)$ is a $d\times d$ dimensional matrix.  %% which represents the gradient of $\ell_n(\theta)$ evaluated at $\theta=\theta 

Note first that under general maximum likelihood theory, $\hat\theta_1,\hat\theta_2$ and $\hat\theta_1,\hat\theta_0$ all converge to $\theta_0^*$ under $(H_0)$, when $n_1\to\infty$ and $n-n_1\to\infty$. 
%Then let $N=(n_1(n-n_1))/n$. 
From Taylor developments and using the fact that $\nabla\ell_n(\hat\theta_0)=0$ we have:
\begin{align}\label{eq:theta}
\tilde\ell_n(\theta_0^*) & = \tilde\ell_n(\hat\theta_0) +\frac 12 (\theta_0^*-\hat\theta_0)^{\top}\nabla^2\ell_n(\theta_0')(\theta_0^*-\hat\theta_0),\nonumber\\
\hat\theta_0-\theta_0^* & =\left(-\nabla^2\ell_n(\theta_0'')\right)^{-1}\nabla\ell_n(\theta_0^*),
\end{align}
where $\theta_0'$ and $\theta_0''$ are on the real line between $\theta_0^*$ and $\hat\theta_0$. From the law of large numbers and the consistency of $\hat\theta_0$ we have that, under $(H_0)$, $-\nabla^2\ell_n(\theta_0')/n$, $-\nabla^2\ell_n(\theta_0'')/n$
%\begin{align*}
%-\frac 1n\nabla^2\ell_n(\theta_0'), \quad -\frac 1n\nabla^2\ell_n(\theta_0'')%\left|\frac 1n\nabla^2\ell_n(\hat\theta_0)-\frac 1n\nabla^2\ell_n(\theta_0')\right|, \quad \left|\frac 1n\nabla^2\ell_n(\hat\theta_0)-\frac 1n\nabla^2\ell_n(\theta_0'')\right|
%\end{align*}
converge towards the Fisher information
\begin{align*}
I(\theta_0^*)=-\mathbb E[\nabla^2\log\left(\mathbb P(X_i;\theta_0^*)\right)].
\end{align*}
%converge towards $0$ in probability, 
From the central limit theorem we have that, under $(H_0)$,$\nabla\ell_n(\theta_0^*)/\sqrt n $ converges toward a centered Gaussian variable in distribution. Using Slutsky's theorem, we directly obtain
\begin{align}\label{eq:1}
\tilde\ell_n(\hat\theta_0) & = \tilde\ell_n(\theta_0^*) +\frac 12\nabla \tilde\ell_n(\theta_0^*)^{\top}\left(-\nabla^2\tilde\ell_n(\theta_0^*)\right)^{-1}\nabla \tilde\ell_n(\theta_0^*)+o_{\mathbb P}(1).
\end{align} 
Then, using the decomposition in the right-hand side of Equation~\eqref{eq:0} for $\nabla \tilde\ell_n(\theta_0^*)$, we have:
\begin{align*}
\tilde\ell_n(\hat\theta_0) & =\tilde\ell_n(\theta_0^*)\\
&\quad+\frac 12 \left[\left(-\frac 1n\sum_{i=1}^{n}\nabla^2 \log\left(\mathbb P(X_i;\theta_0^*)\right)\right)^{-1/2}\!\!\frac 1{\sqrt{n}}\sum_{i=1}^{n_1}\nabla \log\left(\mathbb P(X_i;\theta_0^*)\right)\right]^{\otimes 2}\\
&\quad+\frac 1{2}\left[\left(-\frac 1n\sum_{i=1}^{n}\nabla^2 \log\left(\mathbb P(X_i;\theta_0^*)\right)\right)^{-1/2}\!\!\frac 1{\sqrt{n}}\sum_{i=n_1+1}^{n}\nabla \log\left(\mathbb P(X_i;\theta_0^*)\right)\right]^{\otimes 2}\\
&\quad+\frac 1{\sqrt{n}} \sum_{i=1}^{n_1}\nabla \log\left(\mathbb P(X_i;\theta_0^*)\right)^{\top}\left(-\frac 1n\sum_{i=1}^{n}\nabla^2 \log\left(\mathbb P(X_i;\theta_0^*)\right)\right)^{-1}\!\!\frac 1{\sqrt{n}}\sum_{i=n_1+1}^{n}\nabla \log\left(\mathbb P(X_i;\theta_0^*)\right)\\
&\quad+o_{\mathbb P}(1).
\end{align*}
From the same arguments, we have the following expression of $\ell_n(\hat\theta_1,\hat\theta_2) $:
\begin{align*}
\ell_n(\hat\theta_1,\hat\theta_2) & = \ell_n(\theta_0^*,\theta_0^*) +\frac 12\nabla \ell_n(\theta_0^*,\theta_0^*)^{\top}\left(-\nabla^2\ell_n(\theta_0^*,\theta_0^*)\right)^{-1}\nabla \ell_n(\theta_0^*,\theta_0^*)+o_{\mathbb P}(1)\\
&=\ell_n(\theta_0^*,\theta_0^*)\\
&\quad+\frac 12 \left[ \left(-\frac 1{n_1}\sum_{i=1}^{n_1}\nabla^2 \log\left(\mathbb P(X_i;\theta_0^*)\right)\right)^{-1/2}\!\!\frac 1{\sqrt{n_1}}\sum_{i=1}^{n_1}\nabla \log\left(\mathbb P(X_i;\theta_0^*)\right)\right]^{\otimes 2}\\%\sum_{i=1}^{n_1}\nabla \log\left(\mathbb P(X_i;\theta_0^*)\right)^{\top}
&\quad+\frac 12\left[ \left(-\frac 1{n-n_1}\sum_{i=n_1+1}^{n}\nabla^2 \log\left(\mathbb P(X_i;\theta_0^*)\right)\right)^{-1/2}\!\!\frac 1{\sqrt{n-n_1}}\sum_{i=n_1+1}^{n}\nabla \log\left(\mathbb P(X_i;\theta_0^*)\right)\right]^{\otimes 2}\\
&\quad+o_{\mathbb P}(1).
\end{align*}
From the consistency of
\begin{align*}
\frac 1{n_1}\sum_{i=1}^{n_1}\nabla^2 \log\left(\mathbb P(X_i;\theta_0^*)\right), \quad \frac{1}{n-n_1}\sum_{i=n_1+1}^{n}\nabla^2 \log\left(\mathbb P(X_i;\theta_0^*)\right), \quad\frac 1{n}\sum_{i=1}^{n}\nabla^2 \log\left(\mathbb P(X_i;\theta_0^*)\right)
\end{align*}
towards $I(\theta_0^*)$ we can replace $\sum_{i=1}^{n_1}\nabla^2 \log\left(\mathbb P(X_i;\theta_0^*)\right)/n_1$ and $\sum_{i=n_1+1}^{n}\nabla^2 \log\left(\mathbb P(X_i;\theta_0^*)\right)/(n-n_1)$ by $\hat I(\theta_0^*)$ in the above equation. Taking the difference between $2\ell_n(\hat\theta_1,\hat\theta_2)$ and $2\tilde\ell_n(\hat\theta_0)$ we conclude using the consistency of $\hat\theta$ towards $\theta_0$.% and the continuous mapping theorem.

\subsubsection{Proof of Theorem~\ref{th:likeli_test_small}}
The proofs of 1. and 2. of the theorem are identical, therefore only the proof of 1. is presented.
We first write:
\begin{align*}%\label{eq:2}
\ell_n(\hat\theta_1,\hat\theta_2) & = \sum_{i=1}^{n_1} \log \mathbb P(X_i;\hat{\theta}_1)+\sum_{i=n_1+1}^{n} \log \mathbb P(X_i;\hat{\theta}_2)\nonumber\\
&=\sum_{i=1}^{n_1} \log \mathbb P(X_i;\hat{\theta}_1)+\sum_{i=n_1+1}^{n} \log \mathbb P(X_i;\theta_0^*)\nonumber\\
&\quad+\frac 12\left[\left(-\frac 1n\sum_{i=1}^{n}\nabla^2 \log\left(\mathbb P(X_i;\theta_0^*)\right)\right)^{-1/2}\!\!\frac 1{\sqrt{n-n_1}}\sum_{i=n_1+1}^{n}\nabla \log\left(\mathbb P(X_i;\theta_0^*)\right)\right]^{\otimes 2}\nonumber\\
&\quad+o_{\mathbb P}(1),
\end{align*}
where we used a similar argument as in Equation~\eqref{eq:1} and we replaced $\sum_{i=n_1+1}^{n}\nabla^2 \log\left(\mathbb P(X_i;\theta_0^*)\right)/(n-n_1)$ by $\hat I(\theta_0^*)$. Since $n_1$ is fixed, the sum $\sum_{i=n_1+1}^{n}\nabla \log\left(\mathbb P(X_i;\theta_0^*)\right)$ can be replaced by the sum $\sum_{i=1}^{n}\nabla \log\left(\mathbb P(X_i;\theta_0^*)\right)$ using the fact that $\sum_{i=1}^{n_1}\nabla \log\left(\mathbb P(X_i;\theta_0^*)\right)/\sqrt{n-n_1}$ tends towards $0$ in probability. We finally get:
\begin{align}\label{eq:2}
\ell_n(\hat\theta_1,\hat\theta_2) &=\sum_{i=1}^{n_1} \log \mathbb P(X_i;\hat{\theta}_1)+\sum_{i=n_1+1}^{n} \log \mathbb P(X_i;\theta_0^*)\nonumber\\
&\quad+\frac 12\left[\left(-\frac 1n\sum_{i=1}^{n}\nabla^2 \log\left(\mathbb P(X_i;\theta_0^*)\right)\right)^{-1/2}\!\!\frac 1{\sqrt{n}}\sum_{i=1}^{n}\nabla \log\left(\mathbb P(X_i;\theta_0^*)\right)\right]^{\otimes 2}+o_{\mathbb P}(1).
\end{align}
Next, we write
\begin{align*}
\tilde\ell_n(\hat\theta_0) & =\sum_{i=1}^{n_1} \log \mathbb P(X_i;\hat{\theta}_0)+\sum_{i=n_1+1}^{n} \log \mathbb P(X_i;\hat{\theta}_0)
\end{align*}
and
\begin{align*}
\sum_{i=n_1+1}^{n} \log \mathbb P(X_i;\theta_0^*) & = \sum_{i=n_1+1}^{n} \log \mathbb P(X_i;\hat{\theta}_0) + (\hat\theta_0-\theta^*)^{\top}\sum_{i=n_1+1}^n\nabla \log \mathbb P(X_i;\hat{\theta}_0)\\
&\quad+\frac 12 (\theta_0^*-\hat\theta_0)^{\top}\sum_{i=n_1+1}^n\nabla^2 \log \mathbb P(X_i;\theta'_0),
%\hat\theta_0-\theta_0^* & =\left(-\nabla^2\ell_n(\theta_0'')\right)^{-1}\nabla\ell_n(\theta_0^*),
\end{align*}
where $\theta_0'$ is on the real line between $\theta_0^*$ and $\hat\theta_0$. Since $n_1$ is fixed and $\sum_{i=1}^n\nabla \log \mathbb P(X_i;\hat{\theta}_0)=0$, we have
\begin{align*}
(\hat\theta_0-\theta^*)^{\top}\sum_{i=n_1+1}^n\nabla \log \mathbb P(X_i;\hat{\theta}_0)=-(\hat\theta_0-\theta^*)^{\top}\sum_{i=1}^{n_1}\nabla \log \mathbb P(X_i;\hat{\theta}_0)=o_{\mathbb P}(1).
\end{align*}
From Equation~\eqref{eq:theta} and using the same arguments as in the development of Equation~\eqref{eq:1}, we finally have:

\begin{align*}
\tilde\ell_n(\hat\theta_0) & =\sum_{i=1}^{n_1} \log \mathbb P(X_i;\hat{\theta}_0)+\sum_{i=n_1+1}^{n} \log \mathbb P(X_i;\theta_0^*)- \frac 12 (\theta_0^*-\hat\theta_0)^{\top}\sum_{i=n_1+1}^n\nabla^2 \log \mathbb P(X_i;\theta'_0)\\
&=\sum_{i=1}^{n_1} \log \mathbb P(X_i;\hat{\theta}_0)+\sum_{i=n_1+1}^{n} \log \mathbb P(X_i;\theta_0^*)\\
&\quad + \frac 12 \frac{n-n_1}{n}\left[\left(-\frac 1n\sum_{i=1}^{n}\nabla^2 \log\left(\mathbb P(X_i;\theta_0^*)\right)\right)^{-1/2}\!\!\frac 1{\sqrt{n}}\sum_{i=1}^{n}\nabla \log\left(\mathbb P(X_i;\theta_0^*)\right)\right]^{\otimes 2}+o_{\mathbb P}(1),
\end{align*}
where we replaced $\sum_{i=n_1}^{n}\nabla^2 \log\left(\mathbb P(X_i;\theta_0^*)\right)/(n-n_1)$ by $\sum_{i=1}^{n}\nabla^2 \log\left(\mathbb P(X_i;\theta_0^*)\right)/n$ in the above expression. Taking the difference between Equation~\eqref{eq:2} and the last equation gives the desired result.% we directly obtained the result. 

\newpage
\bibliographystyle{unsrt}
\bibliography{biblio}

\end{document}